\documentclass[twocolumn,superscriptaddress,showpacs,preprintnumbers,amsmath,amssymb,aps,prb]{revtex4}
\usepackage{graphicx}
\usepackage{dcolumn}
\usepackage{bm}
\begin{document}
\title{The fluctuations, under time reversal, of the natural time and the entropy  distinguish similar looking electric signals of different dynamics}
\author{P. A. Varotsos}
\email{pvaro@otenet.gr} \affiliation{Solid State Section and Solid
Earth Physics Institute, Physics Department, University of Athens,
Panepistimiopolis, Zografos 157 84, Athens, Greece}
\author{N. V. Sarlis}
\affiliation{Solid State Section and Solid Earth Physics
Institute, Physics Department, University of Athens,
Panepistimiopolis, Zografos 157 84, Athens, Greece}
\author{E. S. Skordas}
\affiliation{Solid State Section and Solid Earth Physics
Institute, Physics Department, University of Athens,
Panepistimiopolis, Zografos 157 84, Athens, Greece}
\author{M. S. Lazaridou}
\affiliation{Solid State Section and Solid Earth Physics
Institute, Physics Department, University of Athens,
Panepistimiopolis, Zografos 157 84, Athens, Greece}

\begin{abstract}
We show that the scale dependence of the fluctuations
of the natural time itself under time reversal provides a useful
tool for the discrimination of seismic electric signals ({\em critical}
dynamics) from  noises emitted from manmade sources as well as for
the determination of the scaling exponent.
We present recent data of electric signals detected at the Earth's
surface, which confirm  that the value of the entropy in
natural time as well as its value under time reversal are smaller
than that of the entropy of a ``uniform'' distribution.
\end{abstract}
 \maketitle

\section{Introduction}
In a time series comprising $N$ events, the natural time $\chi_{k}
= k/N$ serves as an index\cite{NAT01,NAT02,NAT02A} for the occurrence of
the $k$-th event. In natural time analysis, the time evolution of
the pair of the two quantities ($\chi_k, Q_k$) is considered,
where $Q_k$ denotes in general a quantity proportional to the
energy released during  the $k$-th event. In the case of
dichotomous  electric signals (e.g., seismic electric signal (SES) activities, i.e.,   low frequency
$\leq 1$Hz electric signals that precede earthquakes, e.g., see
Refs.\onlinecite{varbook,nat,tecto88,VAR03b,VAR03b1,VAR03b2,prb,grl,jap,newbook}) $Q_k$ stands for the
duration of the $k$-th pulse (cf. The SES activities should not be confused with pulses of very short durations observed some minutes before earthquakes\cite{pulses}). It has been shown\cite{ABE05} that natural time domain  is optimal for enhancing the signals' localization in the time-frequency space, thus conforming to the desire to reduce uncertainty and extract signal information as much as possible.    The entropy $S$ in natural time is
defined\cite{NAT03B} as the derivative with respect to $q$ of the fluctuation function $\langle \chi^q \rangle-\langle \chi \rangle^q$ at $q=1$:
\begin{equation}
S \equiv  \langle \chi \ln \chi \rangle - \langle \chi \rangle \ln
\langle \chi \rangle \label{Seq}
\end{equation}
where $\langle f( \chi) \rangle = \sum_{k=1}^N p_k f(\chi_k )$ and
$p_k=Q_{k}/\sum_{n=1}^{N}Q_{n}$. It is  dynamic entropy\cite{NAT04,NAT05} and exhibits\cite{NAT05B} concavity, positivity and Lesche\cite{LES82,LES04} stability. Note that $S$ should not be confused with $\text{Cov} [ \chi, \ln \chi] \equiv \langle \chi \ln \chi \rangle-\langle \chi \rangle \langle \ln \chi \rangle $ since in general $\langle \ln \chi \rangle \neq \ln \langle \chi \rangle$.
 The value of the entropy upon
considering the time reversal ${\cal T}$, i.e., ${\cal T}
p_k=p_{N-k+1}$, is labelled by $S_-$.
The value of $S_-$ is\cite{NAT05B,VAR06a,VAR06b}, in general, different from $S$, and thus  $S$ does satisfy the conditions to be ``causal'' in the following
 sense (see Ref.\onlinecite{NAT05B} and references therein): When studying a dynamical system evolving in time, the ``causality'' of an operator
 describing this evolution assures that the values assumed by the operator, at each time instant, depends solely
 on the past values of the system. Hence, a ``causal'' operator should be able to represent the evolution
 of the system according to the (true) time arrow, thus the operator can represent a real physical system evolving in time and reveal the differences arising upon time-reversal of the series.

The statistical properties of $S$ and $S_-$ have been studied in a
variety of models\cite{VAR06a,VAR06b}.  In the case of  a
``uniform'' distribution  $S=S_-=S_u (=\ln 2 /2-1/4\approx
0.0966$). The ``uniform'' distribution (defined in
Refs.\onlinecite{NAT01,NAT03}) has been analytically studied in
Ref.\onlinecite{NAT04} and corresponds to the case when $Q_k$ are
independent and identically distributed (IID) positive random
variables of finite variance including the case of Markovian
dichotomous electric signals studied in Ref.\onlinecite{NAT03B}.
The ``uniform'' distribution corresponds to $p(\chi)=1$, where
$p(\chi )$ is a continuous probability density function (PDF)
corresponding to the point probabilities $p_k$ used so far.  When
$Q_k$ of a ``uniform''  distribution are perturbed by a small
linear trend, we find\cite{VAR06a} (see also Eq.(\ref{bla}),
below) that $(S-S_u)(S_{-}-S_u)<0$ (cf. this simple example, which
shows that  $S$ captures the effect of a linear trend, may be
considered as clarifying the meaning of $S$, see Section V of
Ref.\onlinecite{VAR06a}). Another model studied is when the
increments of $Q_k$ are positive IID, in this case we
find\cite{VAR06b} that $S\approx 0.048$ and $S_-\approx 0.088$
which are both smaller than $S_u$. The same holds, i.e., that both
$S$ and $S_-$ are smaller than $S_u$, in the examples of an on-off
intermittency model discussed in Ref.\onlinecite{VAR06a} as well
as for a multiplicative cascades model\cite{VAR06b} adjusted to
describe turbulence data.

A case of practical importance is that of the SES activities.
SES activities  ({\em critical} dynamics) exhibit infinitely ranged long-range temporal correlations\cite{NAT03,VAR06a,VAR06b} which are
destroyed\cite{VAR06b} after shuffling the durations $Q_k$
randomly. An interesting property emerged from the data analysis
of several SES activities refers to the fact\cite{VAR06a} that
{\em both} $S$ and $S_-$ values are smaller than the value of $S_u$, i.e.,
\begin{equation}
S, S_- < S_u,
\end{equation}
in addition to the fact that for SES activities\cite{NAT01,NAT02A,NAT02} the variance
\begin{equation}
\kappa_1\equiv \langle \chi^2 \rangle -\langle \chi \rangle^2\approx 0.070.
\end{equation}
These findings -which do {\em not} hold\cite{NAT05B} for
``artificial'' noises (AN) (i.e., electric signals emitted from
manmade sources)- have been supported by numerical
simulations in fractional Brownian motion (fBm) time
series\cite{VAR06a,VAR06b} that have an exponent $\alpha_{DFA}$, resulted
from the Detrended Fluctuation Analysis (DFA)\cite{p18,p19}, close
to unity. This model have been applied since  fBm (with a self-similarity index $H\approx 1$) has been found\cite{WER05} as an appropriate type of modeling process for the SES activities. These simulations resulted in values of $S$ and $S_-$  that do obey relation (2) (see Fig.4 of Ref.\onlinecite{VAR06a})  and $\kappa_1\approx 0.070$ (see Fig.3 of Ref.\onlinecite{VAR06b}).   It was then conjectured\cite{VAR06a} that the validity of the relation (2) stems from infinitely ranged long-range temporal correlations (cf. $H \approx 1$). On the other hand, for  short-range temporal correlations (e.g. when modeling $Q_k$ by an autoregressive process $Q_k=a Q_{k-1}+g_k+c, |a|<1$ and $c$ stands for an appropriate constant to ensure positivity of $Q_k$  or $Q_k= |aQ_{k-1}+g_k|$ where $g_k$ is Gaussian IID variables) the values of both $S$ and $S_-$ approach (see Appendix A) that of $S_u$ and $\kappa_1 \rightarrow \kappa_u$, where $\kappa_u=1/12$ denotes the corresponding value of the``uniform'' distribution\cite{VAR06b}.

The scope of this
paper is twofold: First, in Section II, we point out the usefulness of the study of the fluctuations of
the natural time itself under time reversal. In particular, it enables the determination of the scaling exponent, thus allowing the distinction of SES activities from similar looking AN. Second,  in Section III, we provide the most recent experimental
data that strengthen the validity of the relations (2) and (3)  for SES activities.
The earthquakes that followed the latter SES activities are described in Section IV. Section V, summarizes our conclusions.

\section{The fluctuations of natural time under time reversal}
 The way through
which the entropy in natural time captures  the influence of the
effect of a small linear trend has been studied, as mentioned, in
Ref.\onlinecite{VAR06a} on the basis of the parametric family of PDFs:
$p(\chi;\epsilon)=1+\epsilon (\chi -1/2)$, where $\epsilon$ measures the extent of the linear trend. Such a family of PDFs
shares the interesting property ${\cal T}p(\chi;\epsilon)=
p(\chi;-\epsilon)$, i.e, the action of the time reversal is obtained
by simply changing the sign of $\epsilon$. It has been
shown\cite{VAR06a} that the entropy $S(\epsilon )\equiv
S[p(\chi;\epsilon)]$, as well as that of the entropy under time
reversal $S_{-}(\epsilon )\equiv S[{\cal T}p(\chi;\epsilon )]$,
$S_{-}(\epsilon )=S(-\epsilon )$, depend {\em non}-linearly on
the trend parameter $\epsilon$:
\begin{equation} \label{bla}
S(\epsilon)=-\frac{1}{4}+\frac{\epsilon}{72}-\left( \frac{1}{2}+\frac{\epsilon}{12} \right) \ln \left( \frac{1}{2}+\frac{\epsilon}{12} \right).
\end{equation}
However, it would be extremely useful to obtain a {\em linear}
measure of $\epsilon$ in natural time. Actually, this is simply
the average of the natural time itself:
\begin{equation}
\langle \chi \rangle=\int_0^1 \chi p(\chi;\epsilon) d \chi =\frac{1}{2}+\frac{\epsilon}{12}.
\end{equation}
If we consider the fluctuations of this simple measure upon
time-reversal, we can obtain information on the long-range
dependence of $Q_k$. We shall show that a measure of the
long-range dependence emerges in natural time if we study the
dependence of its fluctuations under time-reversal $\Delta \chi_l^2
\equiv  \text{E} [(\langle \chi  \rangle -\langle {\cal T}\chi \rangle )^2 ]$ on the window length
$l$ that is used for the calculation. Since ${\cal
T}p_k=p_{l-k+1}$,  we have
\begin{widetext}
\begin{equation}\label{e1}
  \Delta \chi_l^2 \equiv \text{E} [ ( \langle \chi \rangle  -{ \langle\cal T}\chi \rangle )^2 ] = \text{E} \left\{ \left[ \sum_{k=1}^l \frac{k}{l} \left( p_k-p_{l-k+1} \right) \right]^2 \right\},
  \end{equation}
where the symbol $\text{E}[\ldots]$ denotes the expectation value
obtained when a window of length $l$ is sliding through the time
series $Q_k$. $\text{E}[\ldots]$ is well defined when all the
$\{p_k\}_{k=1,2,\dots l}$ involved in its argument are also well
defined. The evaluation of $\text{E}[\ldots]$ can be carried out
either by full or by Monte Carlo calculation. In order to achieve
this goal, from the original time-series $\{Q_k\}_{k=1,2,\ldots
L}$, we select segments $\{Q_{m_0+i-1} \}_{i=1,2,\ldots l}$ of
length $l$, and the argument of $\text{E}[\ldots]$ is computed by
substituting $p_k=Q_{m_0+k-1}/\sum_{i=1}^lQ_{m_0+i-1}$. The sum of
the resulting values over the number of the selected segments
(different $m_0$) is assigned to $\text{E}[\ldots]$. The full
calculation refers to the case when $m_0$ takes all the $L-l+1
(m_0=1,2,\ldots L-l+1)$ possible values, whereas the Monte Carlo
when $m_0$ is selected randomly.

 By expanding the square in the last part of Eq.(\ref{e1}), we obtain

  \begin{equation}\label{e2}
   \Delta \chi_l^2 = \sum_{k=1}^l \left( \frac{k}{l}\right)^2 \text{E}[ (p_k-p_{l-k+1})^2 ] +
   \sum_{k\neq m} \frac{k m}{l^2} \text{E}[(p_k-p_{l-k+1}) (p_m-p_{l-m+1})].
  \end{equation}
The basic relation\cite{NAT04} that interrelates $p_k$ is
$\sum_{k=1}^l p_k=1$
or equivalently $p_k=1-\sum_{m\neq k} p_m$. By subtracting from
the last expression its value for  $k=l-k+1$, we obtain
$p_k-p_{l-k+1}=-\sum_{m\neq k} (p_m-p_{l-m+1})$, and thus
\begin{equation}\label{e3}
(p_k-p_{l-k+1})^2=-\sum_{m\neq k} (p_k-p_{l-k+1}) (p_m-p_{l-m+1}).
\end{equation}
By substituting Eq.(\ref{e3}) into Eq.(\ref{e2}), we obtain
\begin{equation}\label{e4}
\Delta \chi_l^2 = -\sum_{k=1}^l \left( \frac{k}{l}\right)^2
\sum_{m\neq k} \text{E} [ (p_k-p_{l-k+1}) (p_m-p_{l-m+1}) ] +
\sum_{k\neq m} \frac{k m}{l^2} \text{E}[ (p_k-p_{l-k+1})
(p_m-p_{l-m+1}) ]
\end{equation}
\end{widetext}
which simplifies to
\begin{equation}\label{e5}
\Delta \chi_l^2 = - \sum_{k,m} \frac{(k-m)^2}{l^2} \text{E}[
(p_k-p_{l-k+1}) (p_m-p_{l-m+1}) ]
\end{equation}
The negative sign appears because $(p_k-p_{l-k+1})$ and
$(p_m-p_{l-m+1})$ are in general anti-correlated due to
Eq.(\ref{e3}).  Equation (\ref{e5}) implies that $\Delta \chi_l^2$
measures the long-range correlations in $Q_k$: If we assume that
$-\text{E}[ (p_k-p_{l-k+1}) (p_m-p_{l-m+1})] \propto
(k-m)^{2\chi_H}/l^2$ (cf. $p_k$ scales as $1/l$, e.g. see
\cite{NAT04}), we have that
\begin{equation} \Delta \chi_l^2 \propto
l^{4+2\chi_H}/l^4
\end{equation}
 so that
 \begin{equation}
 \Delta \chi_l (\equiv \sqrt{\Delta \chi_l^2 }) \propto l^{\chi_H},
 \label{scalino}
 \end{equation}
 where
$\chi_H$ is a scaling exponent.

\subsection{Fractional Brownian motion and fractional Gaussian noise time series}
\begin{figure*}
\includegraphics{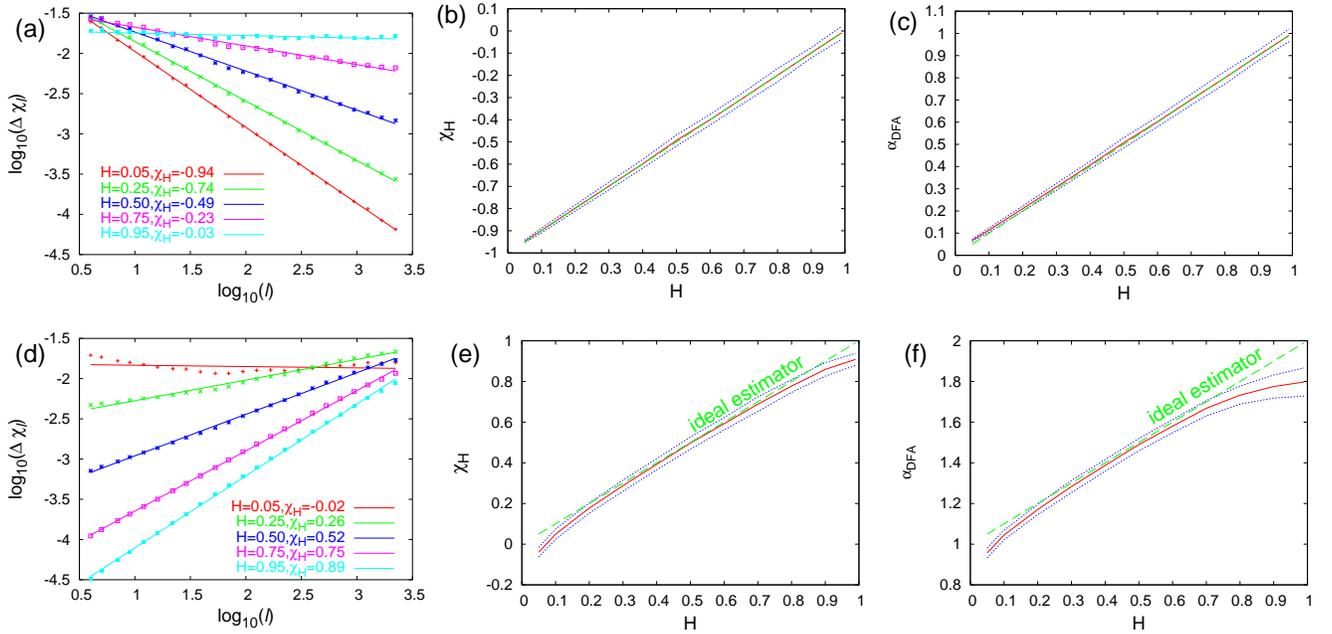}
\caption{\label{mf2} (color online) Examples of log-log plot of
the fluctuations $\Delta \chi_l$ of the natural time under time
reversal versus the scale $l$ for fGn (a) and fBm (d). (b) and (e)
depict the values of the scaling exponent $\chi_H$, introduced
here, versus the true exponent $H$ for fGn and fBm, respectively.
For the sake of comparison, (c) and (f) are similar to (b) and
(e), respectively, but for the DFA exponent $\alpha_{DFA}$. The
(blue) dotted curves show the $\pm \sigma$ deviation from the
average value (obtained after $10^2$ runs) depicted by the (red)
solid curves. The (green) dashed straight lines correspond to the
ideal behavior of each exponent and have been drawn as a guide to
the eye. }
\end{figure*}

In order to examine the validity of the above result
Eq.(\ref{scalino}) when $Q_k$ are coming from fBm or fractional
Gaussian noise (fGn), we employed the following procedure: First,
we generated fBm (or fGn) time-series $X_k$  (consisting of
$2\times10^4$ points) for a given value of $H$ using the
Mandelbrot-Weierstrass function\cite{MAN69,SZU01,inter} as
described in Ref.\onlinecite{VAR06a}. Second, since $Q_k$ should
be positive, we normalized the resulting $X_k$ time-series to zero
mean and unit standard deviation and then added to the normalized
time-series $N_k$ a constant factor $c$ to ensure the positivity
of $Q_k=N_k+c$ (for the purpose of the present study we used
$c=10$). The resulting $Q_k$ time-series were then analyzed and
the fluctuations of $\Delta \chi_l$ versus the scale $l$ are shown
in Figs. \ref{mf2}(a) and \ref{mf2}(d) for fGn and fBm,
respectively. The upper three panels of Fig.\ref{mf2} correspond
to fGn while the lower three to fBm.  We observe (see
Fig.\ref{mf2}(b)) that for fGn  we have the interconnection:
$\chi_H\approx H-1$ corresponding to {\em descending} curves(see
Fig.\ref{mf2}(a)), whereas for fBm the interconnection turns (see
Fig.\ref{mf2}(e)) to: $\chi_H\approx H$ corresponding to {\em
ascending} curves(see Fig.\ref{mf2}(d)).

In order to judge the merits or demerits of the procedure proposed
here for the determination of the scaling exponent, we compare
Figs.\ref{mf2}(b) and \ref{mf2}(e) with  Figs.\ref{mf2}(c) and
\ref{mf2}(f), respectively, that have been obtained by the
well-established DFA method\cite{p18,p19}. This comparison reveals
that  the results are more or less comparable for fGn, while for
fBm  the exponent $\chi_H$ deviates less  from the  behavior of an
ideal estimator of the true scaling exponent (drawn in dashed
green)  compared to $\alpha_{DFA}$, especially for the largest $H$
values.

\subsection{The fluctuations of the natural time to distinguish seismic electric signal activities from similar looking AN}
\begin{figure*}
\includegraphics{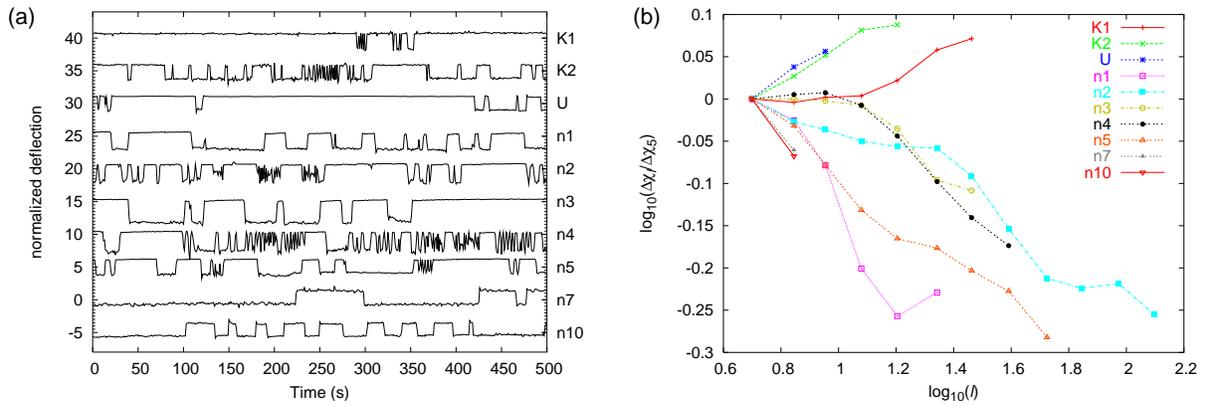}
\caption{\label{mf3} (color online) The log-log plot of $\Delta
\chi_l$ versus the scale $l$ for three SES activities (K1, K2 and
U) and seven AN (n1-n5, n7 and n10) treated in
Ref.\onlinecite{NAT05B} (cf. these signals have enough number of
pulses in order to apply the present analysis). The values of
$\Delta \chi_l$ are divided by the corresponding values $\Delta
\chi _5$ at the scale $l=5$.}
\end{figure*}

 The physical meaning of
the present analysis was further investigated by performing the
same procedure in the time-series of the durations of those
signals analyzed in Ref.\onlinecite{NAT05B} that have enough
number of pulses e.g.$\approx 10^2$ (cf. the signals depicted in
Fig.\ref{mf1} could not be analyzed in view of the  small number
of pulses). The relevant results are shown in Fig.\ref{mf3}. Their
inspection interestingly indicates that all seven AN correspond to
descending $\Delta \chi_l$ curves versus the scale $l$, while the
three SES activities to ascending curves  (in a similar fashion as
in Figs.\ref{mf2}(a) and \ref{mf2}(d), respectively) as expected
from the fact that the latter exhibit\cite{NAT03} infinitely
ranged long-ranged temporal correlations (having $H$ close to
unity), while the former do not. Hence, the method proposed here
enables the detection of long-range correlations even for datasets
of small  size ($\approx 10^2$), thus allowing the distinction of
SES activities from AN.

\section{Recent data of Seismic Electric Signals activities}
First, Fig.\ref{mf1}(a) depicts an electric signal, consisting of a number of
pulses, that has been recorded on November 14, 2006 at a station
labelled\cite{EPAPS74} PIR lying in western Greece (close to Pirgos
city). This signal has been clearly collected at eleven measuring
electric dipoles with electrodes installed at sites that are
depicted in a map given in Ref.\onlinecite{EPAPS74}. The signal is
presented (continuous line in red) in Fig. \ref{mf1}(a) in normalized
units, i.e., by subtracting the mean value and dividing by the
standard deviation.
For the reader's convenience, the corresponding dichotomous
representation is also drawn in Fig. \ref{mf1}(a) with a dotted (blue)
line, while in Fig. \ref{mf1}(c) we show (in red crosses) how the signal
is read in natural time. The computation of $S$ and $S_-$ leads to
the following values: $S=0.070\pm0.012$, $S_-=0.051\pm0.010$. As
for the variance $\kappa_1$, the resulting
value is $\kappa_1=0.062\pm0.010$. These values more or less obey
the conditions (2) and (3) that have
been found to hold for other SES activities\cite{VAR06a}.  Note that the feature of this SES activity, it is similar to the one
observed at the same station before the magnitude $M\approx 6.7$
earthquake that occurred on Jan 8, 2006, see Ref.\onlinecite{VAR06}.

A closer inspection of Fig. \ref{mf1}(a)  reveals the following
experimental fact: An additional electric signal has been also
detected (in the gray shaded area of Fig. \ref{mf1}(a)), which consists of
pulses with markedly smaller amplitude than those of the SES
activity discussed in the previous paragraph. This is reproduced
(continuous line in red) in Fig. \ref{mf1}(b) in an expanded time scale
and for the sake of the reader's convenience its dichotomous
representation is also marked by the dotted (blue) line, which
leads to the natural time representation shown (dotted blue) in
Fig. \ref{mf1}(c). The computation of $S$ and $S_-$ gives
$S=0.077\pm0.004$, $S_-=0.082\pm0.004$, while $\kappa_1$ is found
to be $\kappa_1=0.076\pm0.005$. Hence, these values also obey the
conditions (2) and (3)  for the  classification of this signal as an SES activity.

The two aforementioned signals have been followed by two
significant earthquakes as described in Section IV. This conforms
to their classification as SES activities, which has been
completed in an early version of this paper\cite{ARXIV06} on
November 16, 2006.

Second, very recently, i.e., on July 2, 2007 and July 10,2007, two
separate electric signals were recorded at a station labelled PAT
lying in central Greece (close to Patras city) at
38.32$^o$N21.90$^o$E. The signals are presented (continuous line
in red) in Fig. \ref{mf1} (d) and  \ref{mf1} (e) in normalized
units in a similar fashion as in Figs.\ref{mf1}(a),(b). Their
corresponding dichotomous representation are also drawn with
dotted (blue) lines, while in Fig.\ref{mf1}(f) we show (in red
crosses and blue asterisks, respectively) how the signals are read
in natural time. The computation of $\kappa_1$, $S$ and $S_-$
leads to the  following values: For the signal on July 2, 2007:
$\kappa_1=0.072\pm0.005$, $S=0.073\pm0.007$, $S_-=0.081\pm0.006$,
for the signal on July 10, 2007: $\kappa_1=0.073\pm0.004$,
$S=0.085\pm0.005$, $S_-=0.080\pm0.004$. An inspection of these
values reveals that they obey the conditions (2) and (3) and hence
both signals can be classified as SES activities. The procedure
for the current study of the subsequent seismicity that occurred
after these SES activities is  described in the next Section.

\section{The seismic activity that followed the SES activities}

We discriminate that during the last decade SES activities are
publicized {\em only} when their amplitude indicates that the
impending earthquake has an expected\cite{newbook,VAR06b}
magnitude comparable to 6.0 unit or larger.

\subsection{The case of the SES activities of Figs.3(a),(b)}
\begin{figure*}
\includegraphics{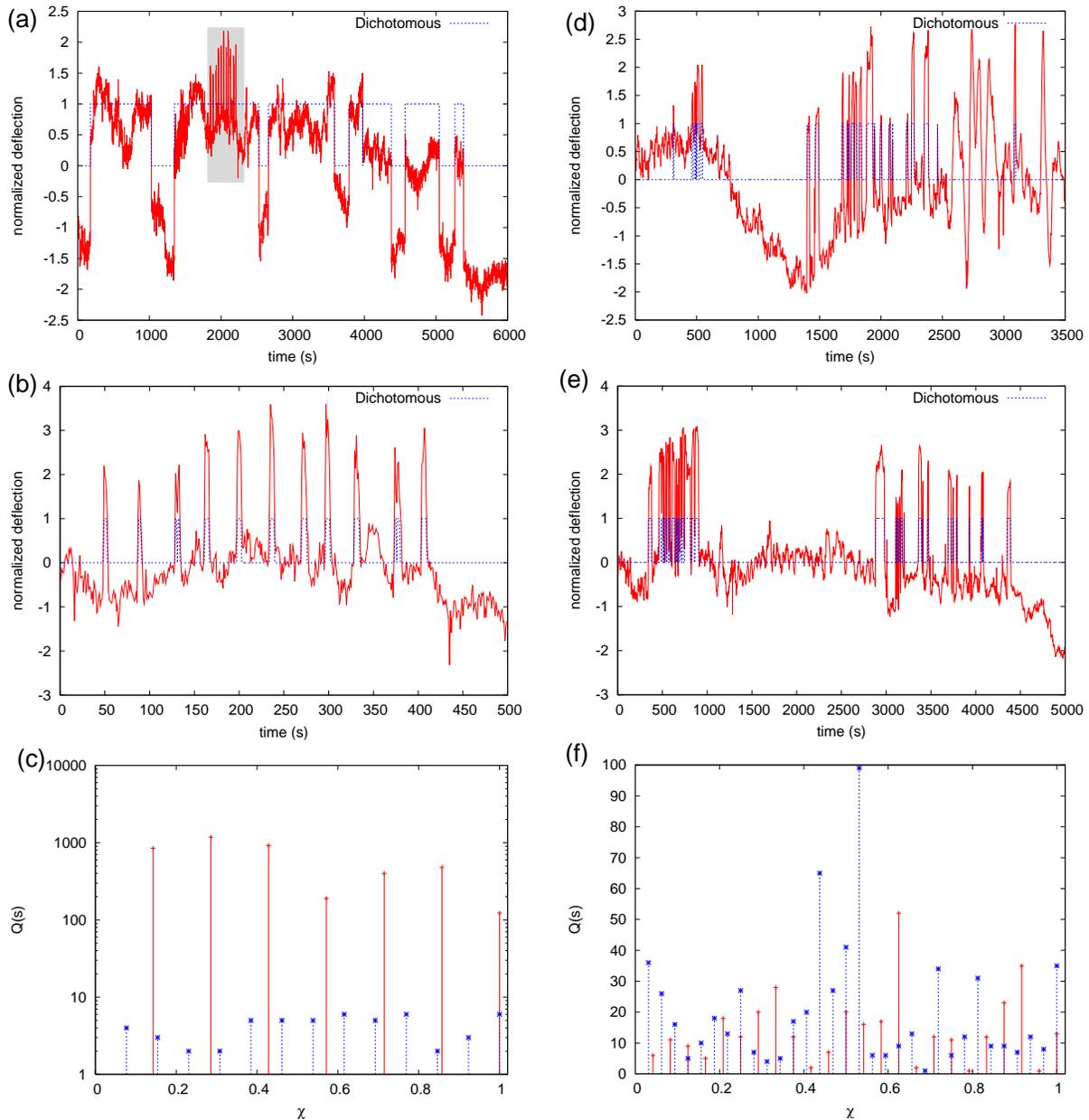}
\caption{\label{mf1} (color online) (a) The electric signal
recorded on November 14, 2006 at PIR station (sampling rate
$f_{exp}=1$Hz). The signal is presented here in normalized units
(see the text). The corresponding dichotomous representation is
shown with the dotted (blue) line. The gray shaded area shows an
additional signal (consisting of pulses of smaller duration)
superimposed on the previously mentioned signal. (b) The signal
belonging to the gray shaded area in (a) is given here in an
expanded time scale, while its dichotomous representation is
marked by the dotted (blue) line. (c) How the signals in (a) and
(b) are read in natural time: the continuous (red crosses) and
dotted (blue asterisks) bars correspond to the durations of the
dichotomous representations marked in (a) and (b), respectively.}
\end{figure*}

According to the Athens observatory (the data of which will be
used here),  a strong earthquake (EQ) with magnitude 5.8-units
occurred at 13:43 UT on February 3, 2007, with epicenter at
$35.8^o$N $22.6^o$E, i.e., almost 80 km to the southwest of the
6.9 EQ of January 8, 2006, (cf. the magnitude announced from
Athens observatory is equal to ML+0.5, where ML stands for the
local magnitude). This was preceded by a 5.2-units EQ that
occurred at 22:25 UT on January 18, 2007 at $34.8^o$N $22.7^o$E.
The occurrence
 of these two EQs confirm the classification
as SES activities of the signals depicted in Figs.\ref{mf1}(a)and
\ref{mf1}(b). (Note that preseismic information based on SES
activities is issued {\em only} when the magnitude of the
strongest EQ of the impending EQ activity is estimated  -by means
of the SES amplitude\cite{EPAPS2}- to be comparable to 6.0 units
or larger\cite{newbook}.)

\begin{figure*}
\includegraphics{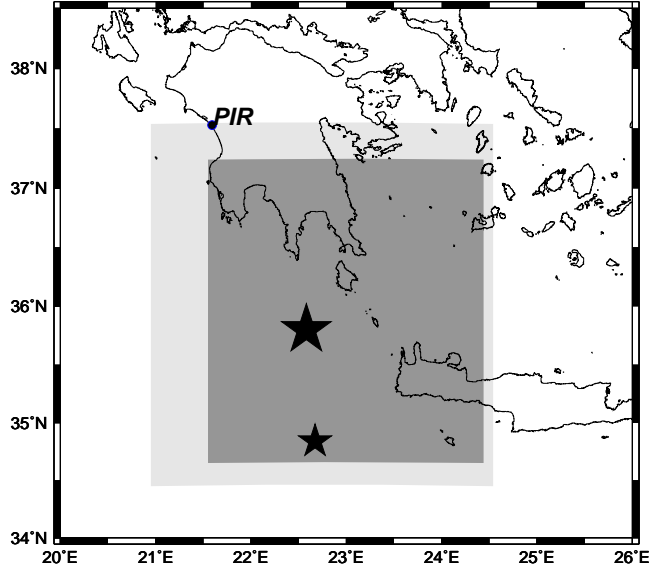}
\caption{Map of the area surrounding the station PIR (solid dot)
and the epicenters of the EQs (stars) with magnitude 5.2-units and
5.8-units that occurred on January 18 and February 3, 2007,
respectively. The seismicity subsequent to the SES activities on
November 14, 2006, has been studied in the gray shaded areas (the
large and the small area are designated A and B, respectively).}
\label{f1}
\end{figure*}
\begin{figure*}
\includegraphics{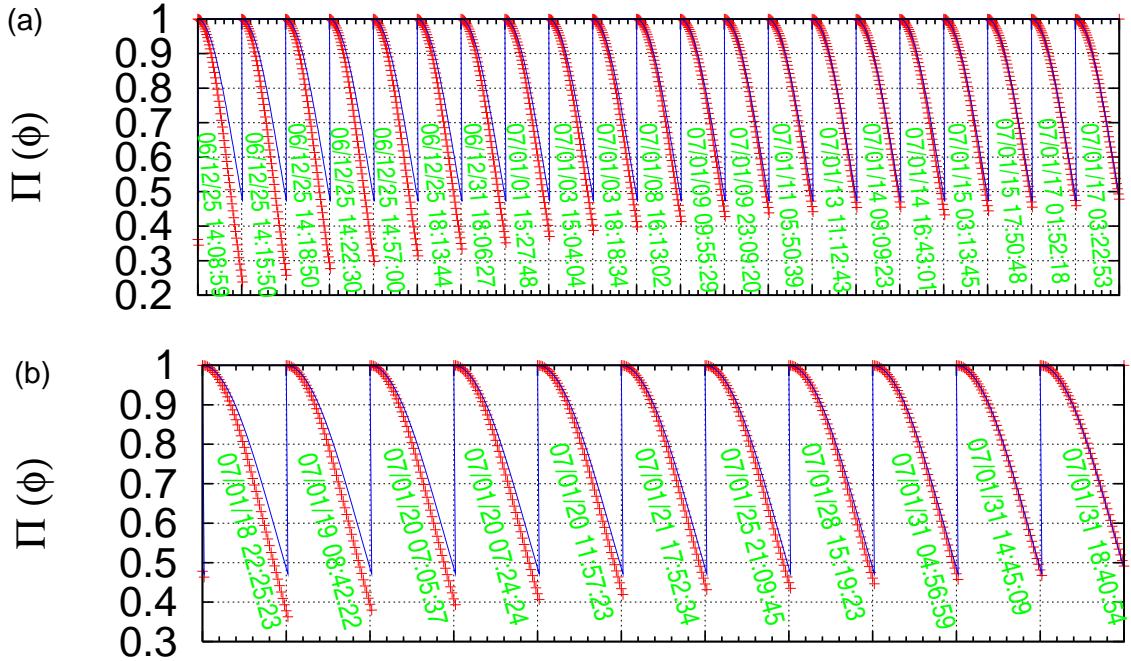}
\caption{(color online) The normalized power spectrum(red) $\Pi
(\phi )$ of the seismicity   as it evolves event by event (whose
date and time of occurrence are written in green in each panel)
after the initiation of the SES activity on November 14, 2006.
The two excerpts presented here, refer to the periods: (a)
December 25, 2006, to January 17, 2007, and (b): January 18  to
January 31, 2007, and correspond to the area B with
$M_{thres}=3.4$.  In each panel only the spectrum in the range
$\phi \in [0,0.5]$ (for the reasons discussed in
Refs.\onlinecite{NAT01,tan05}) is depicted (separated by the
vertical lines), whereas the  $\Pi (\phi )$ of Eq.(\ref{fasma}) is
depicted by blue color. The minor horizontal ticks for $\phi$ are
marked every 0.1.} \label{FigFasma}
\end{figure*}
\begin{figure*}
\includegraphics{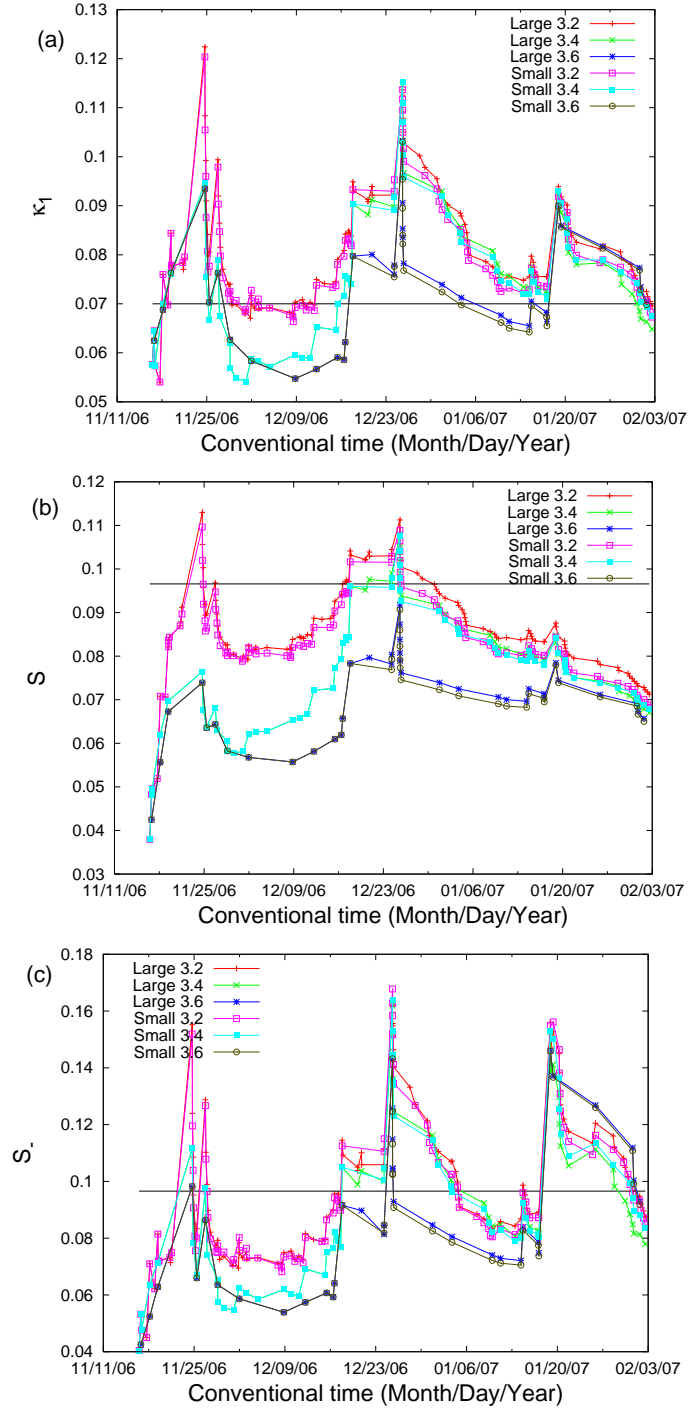}
\caption{(color online) Evolution of the quantities  $\kappa_1$,
$S$ and $S_{-}$; they are shown in (a), (b) and (c), respectively,
for three magnitude thresholds, i.e., $M \geq 3.2, 3.4$ and $M
\geq 3.6$, for both areas, i.e., the large (area A) and the small
(area B). The solid horizontal lines correspond to $\kappa_1=0.07$
and $S_u=\ln 2 /2 -1/4$. } \label{k1Sb}
\end{figure*}

Here, we show that the occurrence times of the aforementioned two
EQs can be estimated by following the procedure described in
Refs.\onlinecite{NAT01,tan05,VAR06a,VAR06b} and using the order
parameter of seismicity proposed in Ref.\onlinecite{tan05}, i.e,
the normalized power spectrum in natural time  $\Pi (\phi )$ as $
\phi \rightarrow 0$ (see also below).   We study how the
seismicity evolved after the recording of the SES activities on
November 14, 2006, at PIR station (which were classified as SES
activities in the initially submitted version of the present paper
on November 16, 2006). The study is made either in  the area
A:N$_{34.45}^{37.54}$E$_{20.95}^{24.54}$ or in  the area
B:N$_{34.65}^{37.24}$E$_{21.55}^{24.44}$ (see Fig.\ref{f1}), by
considering three magnitude thresholds $M_{thres}=3.2$, 3.4 and
3.6 (hence six combinations were studied in total).
 If we set the natural time for
seismicity zero at the initiation of the SES activity at 17:19 UT
on November 14, 2006, we form time series of seismic
events\cite{magest} in natural time for various time windows as
the number $N$ of consecutive (small) EQs increases. We then
compute the normalized power spectrum
\cite{NAT01,tan05,VAR06a,VAR06b} in natural time $\Pi (\phi )$ for
each of the time windows. Excerpts of these results, which refer
to the values during the periods: (a) December 25, 2006, to
January 17, 2007, and (b): January 18 to January 31, 2007 are
depicted with red crosses in Fig.\ref{FigFasma}, respectively.
This figure corresponds to the small area B with $M_{thres}=3.4$.
In the same figure, we plot in blue the normalized power spectrum
obeying the relation\cite{NAT01,NAT02,NAT02A,tan05}
\begin{equation}
\Pi ( \omega ) = \frac{18}{5 \omega^2}
-\frac{6 \cos \omega}{5 \omega^2}
-\frac{12 \sin \omega}{5 \omega^3}
\label{fasma}
\end{equation}
which holds when the system enters the {\em critical} stage
($\omega = 2\pi \phi$, where $\phi$ stands for the natural
frequency\cite{NAT01,NAT02,NAT02A,newbook}). The date and the time
of the occurrence of each small earthquake (with $M \geq 3.4$) that
occurred in the area B, is written in green in each panel (see also
Table \ref{tab60}). An inspection of Fig.\ref{FigFasma}(a) reveals
that the red line approaches the blue line as $N$ increases and a
{\em coincidence} occurs at the small event of magnitude 3.7 that
occurred at 03:22 UT on January 17, 2007, i.e., roughly two days
before the 5.2-units EQ at 22:25 UT on January 18,2007. A similar
behavior is observed in Fig.\ref{FigFasma}(b) in which we see that
a {\em coincidence} occurs at the small event of magnitude 3.6 at
18:40 UT on January 31, 2007, i.e., roughly three days before the
strong EQ of magnitude 5.8-units that occurred at 13:43 UT on February 3, 2007. To
ensure that these two coincidences in Figs.\ref{FigFasma}(a) and
(b) are {\em true} ones\cite{NAT01,tan05,NAT02A,newbook,EPAPS74} (see also
below) we also calculate the evolution of the quantities
$\kappa_1$,$S$ and $S_{-}$   and the results are depicted in
Fig.\ref{k1Sb} for the three magnitude thresholds for
each of the aforementioned two areas A and B.

\begin{figure*}
\includegraphics{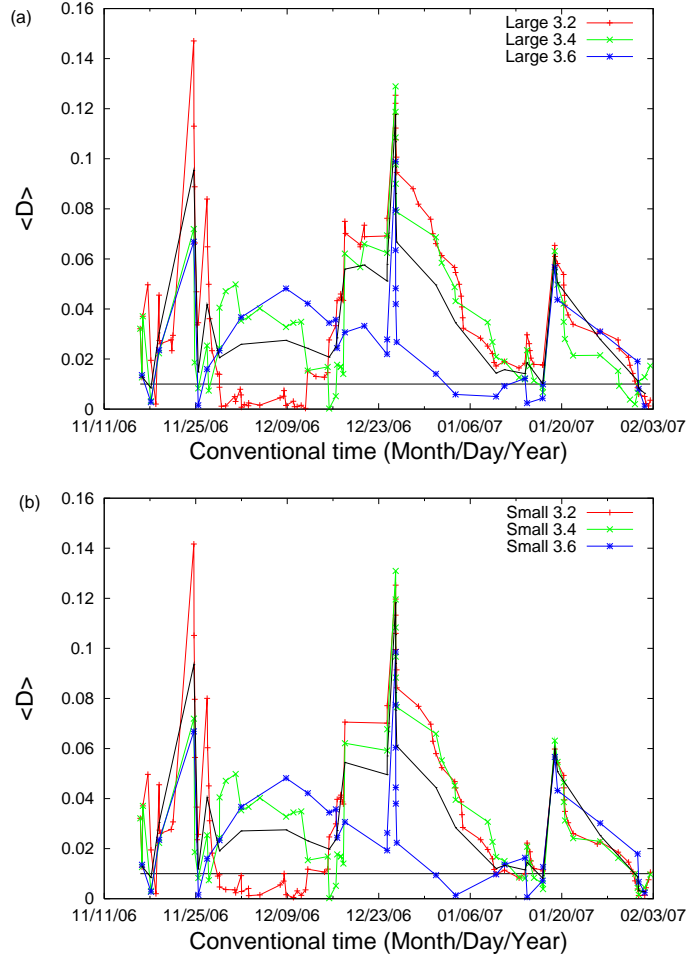}
\caption{(color online)The average distance $\langle D \rangle$
between the calculated and the theoretical $\Pi(\phi )$ curves
versus the conventional time. The calculation of $\langle D
\rangle$ is made upon the occurrence of every consecutive
earthquake when starting the calculation after the initiation of
the SES activities on November 14, 2006 (depicted in
Figs.\ref{mf1}(a),(b)) for  the large (area A) and the small (area
B)  for the three magnitude thresholds $M_{thres}=3.2$, 3.4 and
3.6.  The black solid line corresponds to the mean value obtained
from the three thresholds.} \label{Distance}
\end{figure*}

The conditions for a coincidence to be considered as {\em true}
are the following (e.g., see Ref.\onlinecite{NAT01}, see also
\cite{tan05,NAT02A,newbook,EPAPS74}): First, the `average'
distance $\langle D \rangle$ between the empirical and the
theoretical $\Pi(\phi )$(i.e., the red and the blue line,
respectively, in Fig.\ref{FigFasma}) should
be\cite{NAT01,tan05,NAT02,newbook,EPAPS74} smaller than or equal
to $10^{-2}$. See Fig.\ref{Distance}, where we plot $\langle D
\rangle$ versus the conventional time during the whole period
after the recording of the SES activities on November 14, 2006,
for both areas, i.e., the large one (area A) and the small (area
B) and the three magnitude thresholds. For the sake of the readers
convenience, the mean value of the results obtained for the three
thresholds is also shown in black.  Second, in the examples
observed to date\cite{NAT01,tan05,NAT02A,newbook,EPAPS2,EPAPS74},
a few events {\em before} the coincidence leading to the strong
EQ, the evolving $\Pi(\phi )$ has been found to approach that of
Eq.(1), i.e., the blue one in Fig.\ref{FigFasma} , from {\em
below} (cf. this reflects that during this approach the
$\kappa_1$-value decreases as the number of events increases). In
addition, both values $S$ and $S_{-}$ should be smaller than $S_u$
at the coincidence. Finally, since the process concerned is
self-similar ({\em critical} dynamics), the time of the occurrence
of the (true) coincidence should {\em not} change, in principle,
upon changing either the (surrounding) area or the magnitude
threshold used in the calculation. Note that in
Fig.\ref{Distance}, at the last small events ,i.e., the rightmost
in Figs.\ref{FigFasma}(a) and \ref{FigFasma}(b), respectively
(i.e., the magnitude 3.7 event on January 17, 2007 and the second
event of magnitude 3.6 on January 31, 2007) just before the
occurrences of the 5.2-units and 5.8-units EQs,  in both areas A
and B, the mean value (see the black thick lines in
Fig.\ref{Distance}) of $\langle D \rangle$ obtained from the three
magnitude thresholds become smaller than or equal to   $10^{-2}$ .
Hence, these two coincidences can be considered as {\em true}.

\newpage
\begin{figure*}
\includegraphics{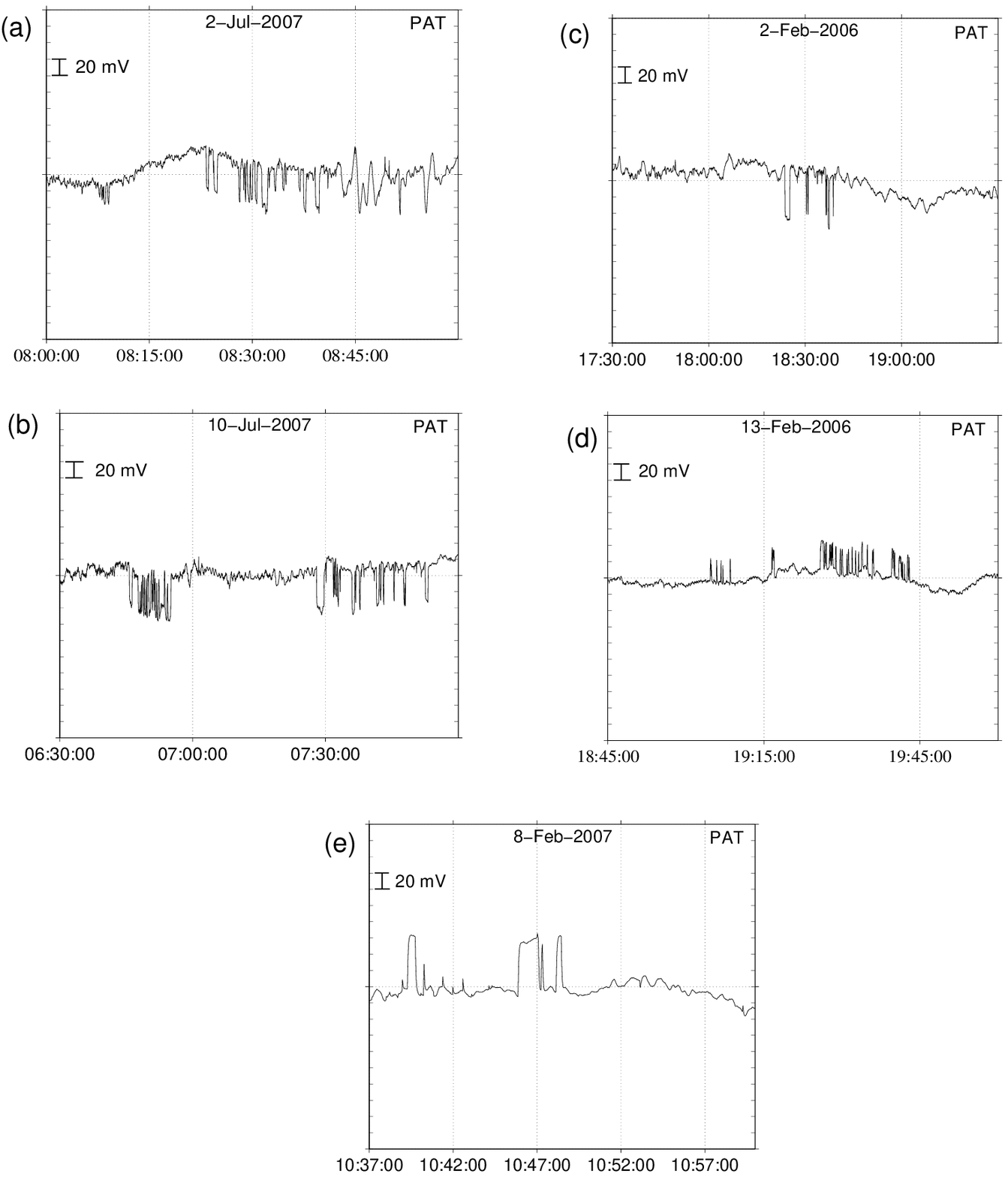}
\caption{\label{Neoreo} Recent SES activities recorded at PAT.}
\end{figure*}
\newpage

In summary, the SES activities recorded on November 14, 2006, at
PIR station (presented in Figs.\ref{mf1}(a),(b)) have been
followed by two EQs with magnitudes 5.2-units and 5.8-units that
occurred on January 18 and February 3, 2007. The time of the
occurrences of these two EQs are determined within a narrow range
of a few days upon analyzing, in natural time, the seismicity
subsequent to the SES activities.

\subsection{The case of the SES activities of Figs.3(d),(e)}
The actual amplitude (in mV) of the most recent SES activities
recorded at PAT on July 2, 2007 and July 10, 2007 (see
Fig.\ref{mf1}(d) and (e), respectively) can be visualized in
Figs.\ref{Neoreo}(a) and  \ref{Neoreo}(b) where the original
recordings of a measuring electric dipole (with length
$L\approx$5km) are reproduced. For the sake of comparison, in
Figs.\ref{Neoreo}(c),(d) we also present the corresponding SES
activities at the same station, i.e., PAT, that
preceded\cite{VAR06b,EPAPS74} the magnitude 6.0-class earthquakes
that occurred with epicenters at 37.6$^o$N20.9$^o$E on April 11
and 12,2006. Furthermore, in Fig.\ref{Neoreo}(e), we show the SES
activity\cite{ARXIV07} at PAT on February 8, 2007, which was
followed by a magnitude class 6.0 earthquake at 38.3$^o$N20.4$^o$E
that occurred on March 25, 2007.

\begin{figure*}
\includegraphics{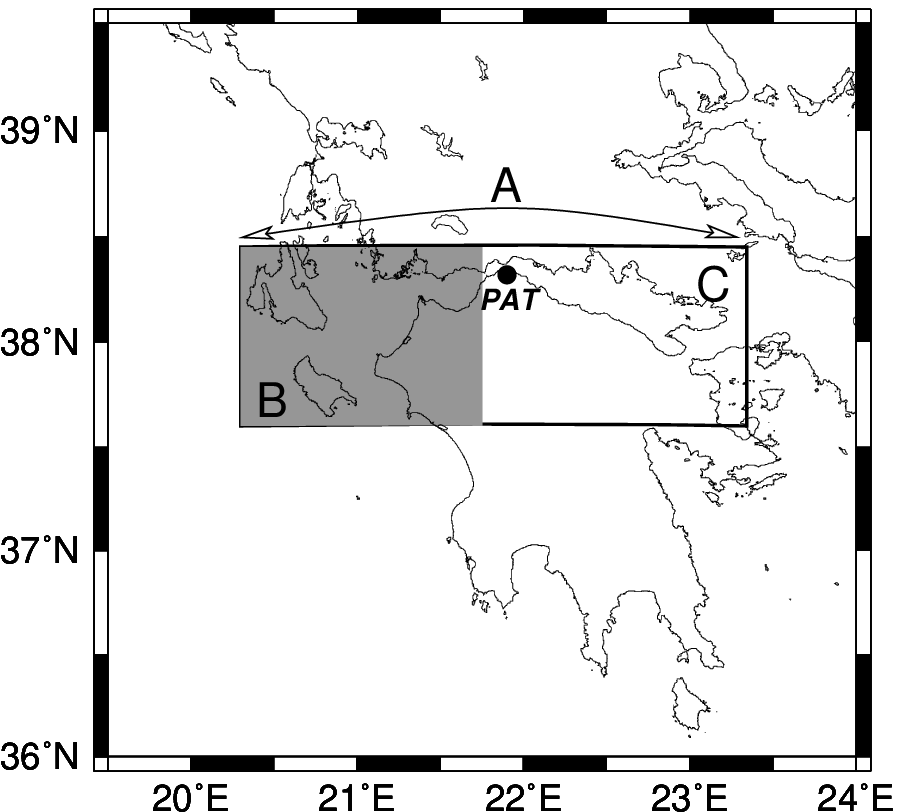}
\caption{\label{Neomap} A map showing the areas A, B and C at
which the analysis of the seismicity in natural time is currently
carried out.}
\end{figure*}

In order to determine the occurrence time of the impending EQs, we
currently apply the procedure explained in the previous subsection
by studying the seismicity in the areas A, B, C  (see
Fig.\ref{Neomap}). Since the result should exhibit {\em spatial}
scale invariance, the epicenter(s) will lie either in the area B
or in C depending on whether the areas A and B or A and C show
{\em true} coincidence.
\newpage
\begin{figure*}
\includegraphics{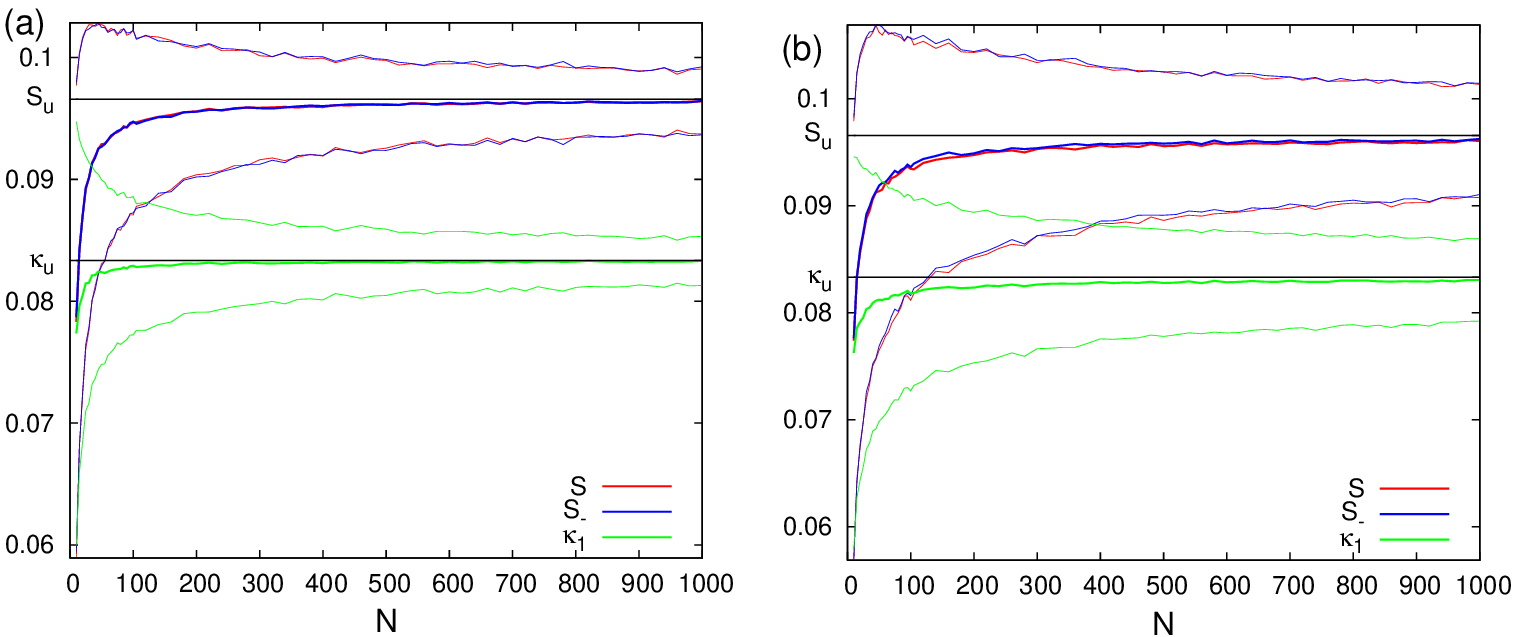}
\caption{(color online)  \label{lug1}The thick colored lines in
(a) and (b) depict the average value  of $S$ (red),$S_-$ (blue)
and $\kappa_1$ (green) versus the number of $Q_k$ for the two
examples (i) and (ii), respectively,  mentioned in the Appendix.
The thinner colored lines refer to the $\pm \sigma$ deviation from
the average value.   }
\end{figure*}

\section{Conclusions}
First, the scale dependence of the fluctuations of the natural
time under time reversal distinguish similar looking electric
signals emitted from systems of different dynamics providing a
useful tool for the determination of the scaling exponent.  In
particular, SES activities ({\em critical } dynamics) are
distinguished from noises emitted from man-made electrical
sources.

Second, recent data of SES activities are presented which  confirm
that the value of the entropy in natural time as well as its value
under time reversal are smaller than that of a ``uniform''
distribution.

\appendix

\section{The case of signals that exhibit short-range temporal correlations}
Here, we present results of modeling $Q_k$ by short-ranged
temporal correlated time-series. Two examples were treated by
numerical simulation: (i) A stationary autoregressive process
$Q_k=a Q_{k-1}+g_k+c$, $|a|<1$, where $g_k$ are Gaussian IID
random variables, and $c$ stands for an appropriate constant to
ensure positivity of $Q_k$.  (ii) $Q_k= |aQ_{k-1}+g_k|$. Figure
\ref{lug1}(a) depicts the results for $S$,$S_-$ and $\kappa_1$ for
the first example versus the number of $Q_k$, whereas
Fig.\ref{lug1}(b) refers to the second example. In both cases $S$
and $S_-$  converge to $S_u$ whereas $\kappa_1$ to the value
$\kappa_u=1/12$ corresponding to the ``uniform'' distribution. For
the reader's convenience, the values of $S_u$ and $\kappa_u$ are
designated by the horizontal solid black lines.

\section{The seismic activity that followed the recent SES
activities of Figs.8(a),(b)}

\begin{turnpage}
\begin{figure}
\includegraphics{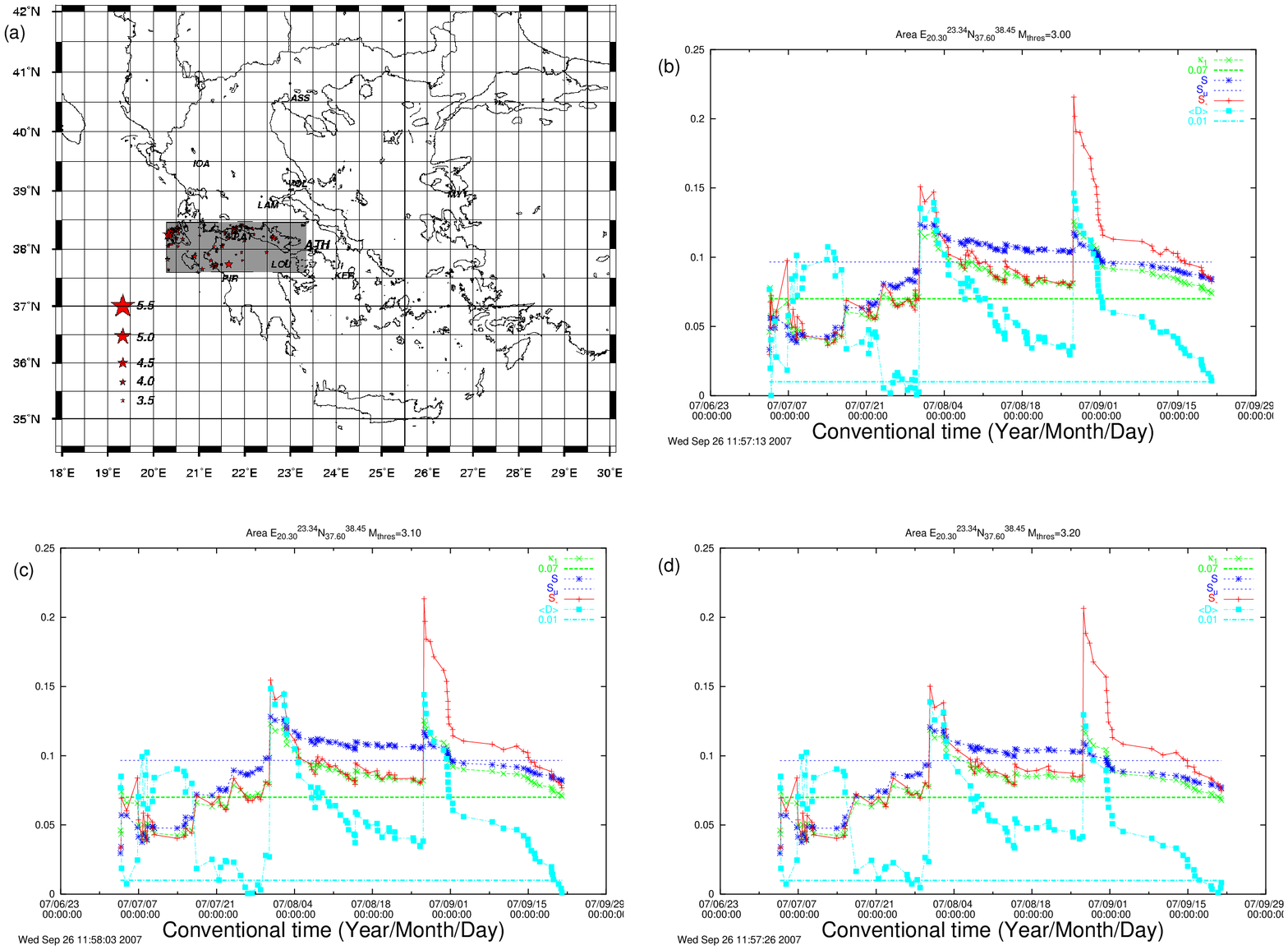}
\caption{(color online) Results of the analysis of seismicity in
the dark gray shaded area of (a) after the SES initiation on July
2, 2007. } \label{A}
\end{figure}
\end{turnpage}

\begin{turnpage}

\begin{figure}
\includegraphics{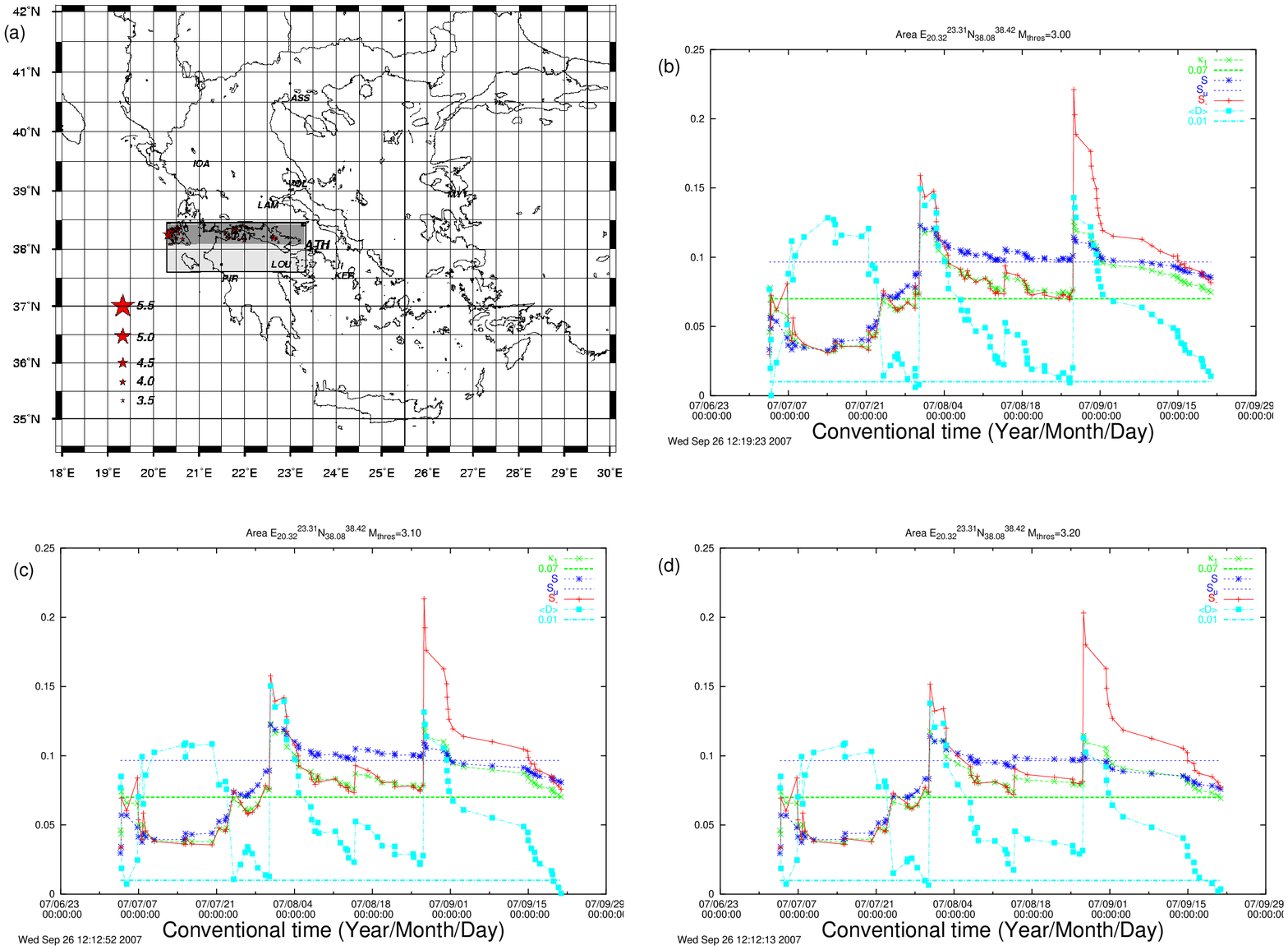}
\caption{(color online) Results of the analysis of seismicity in
the dark gray shaded area of (a) after the SES initiation on July
2, 2007. } \label{S1}
\end{figure}
\end{turnpage}

\begin{turnpage}

\begin{figure}
\includegraphics{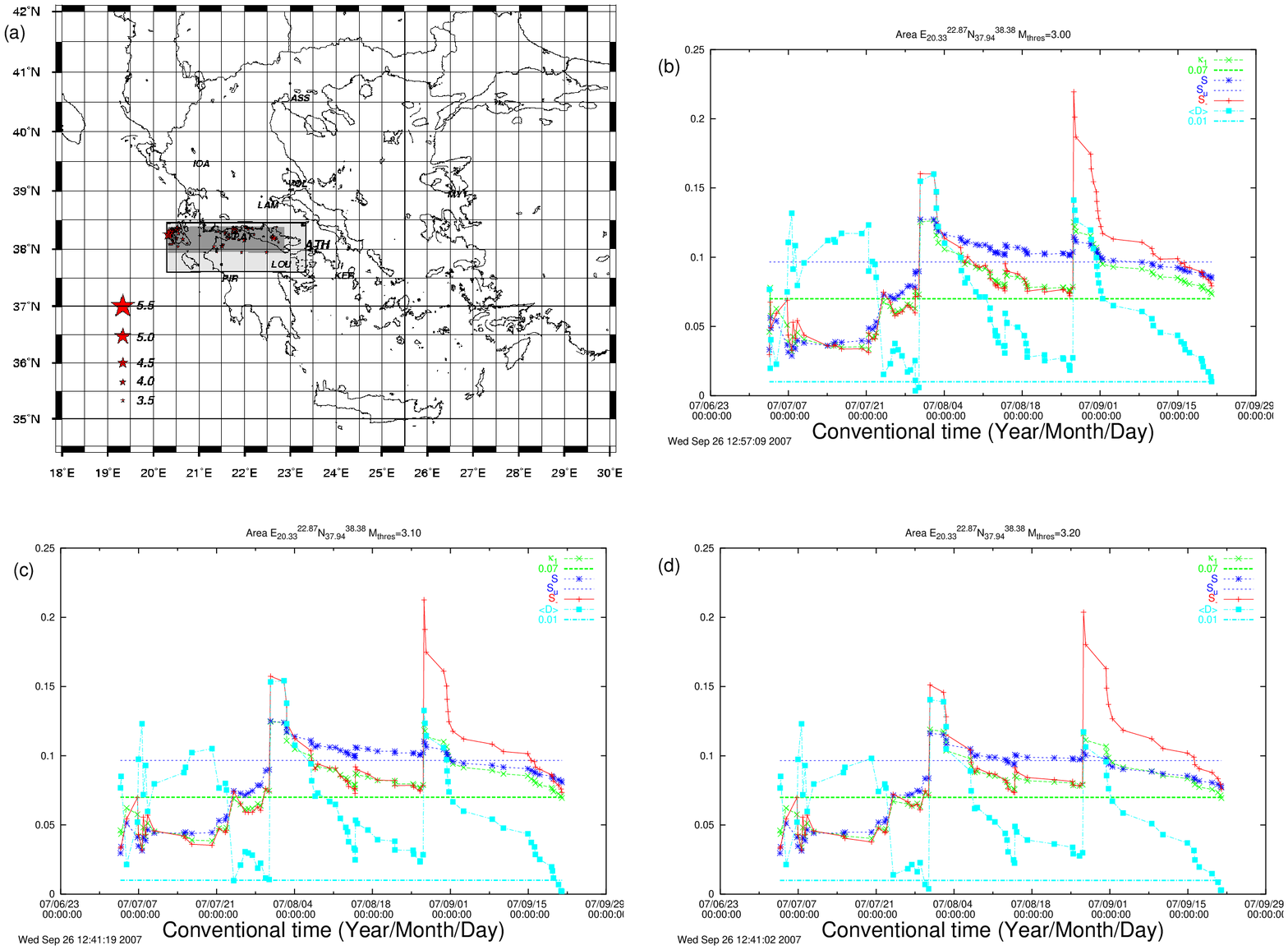}
\caption{(color online) Results of the analysis of seismicity in
the dark gray shaded area of (a) after the SES initiation on July
2, 2007. } \label{S2}
\end{figure}

\end{turnpage}

\begin{turnpage}
\begin{figure}
\includegraphics{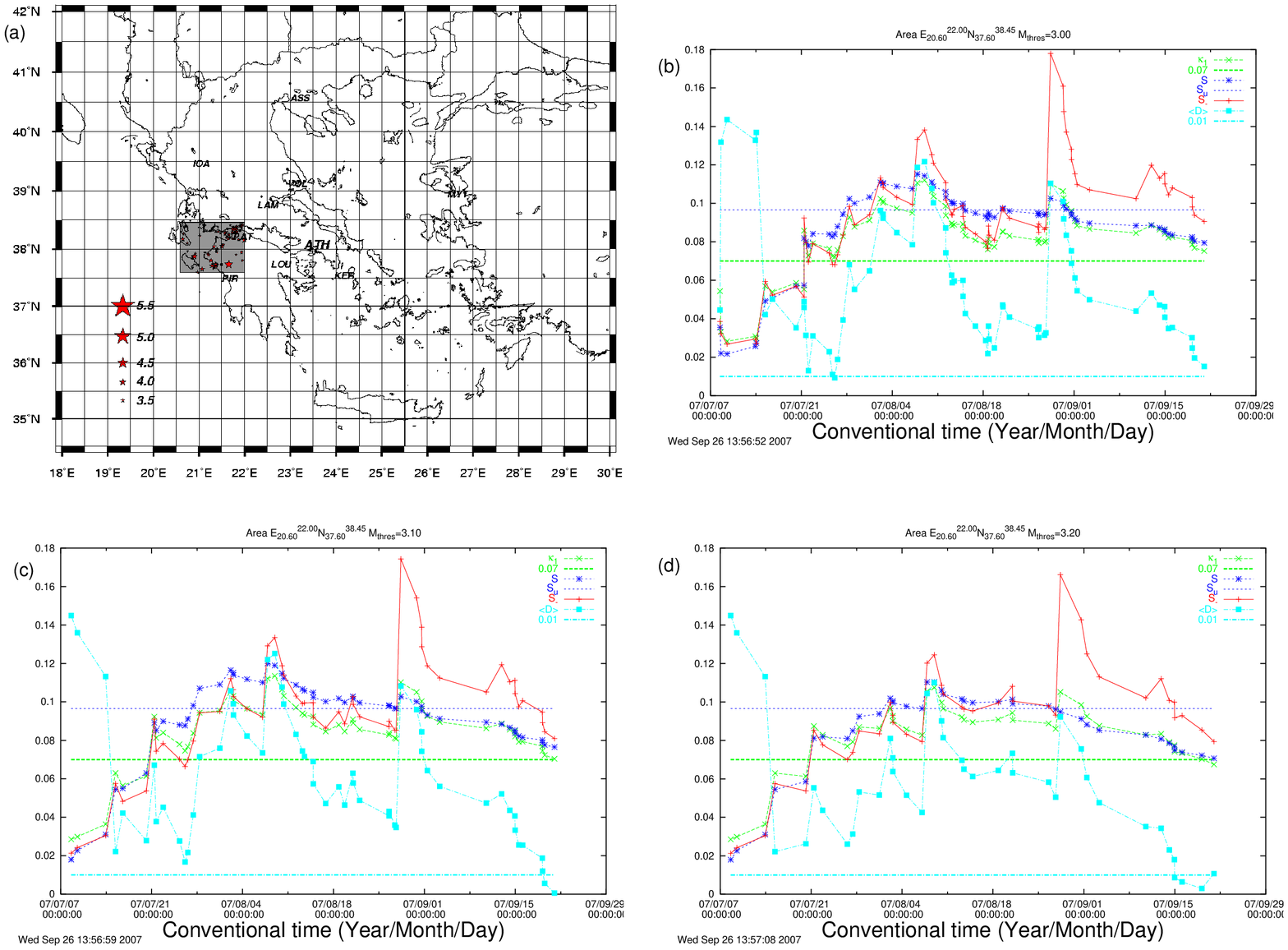}
\caption{(color online) Results of the analysis of seismicity in
the dark gray shaded area of (a) after the SES initiation on July
2, 2007. } \label{S3}
\end{figure}

\end{turnpage}

\begin{turnpage}
\begin{figure}
\includegraphics{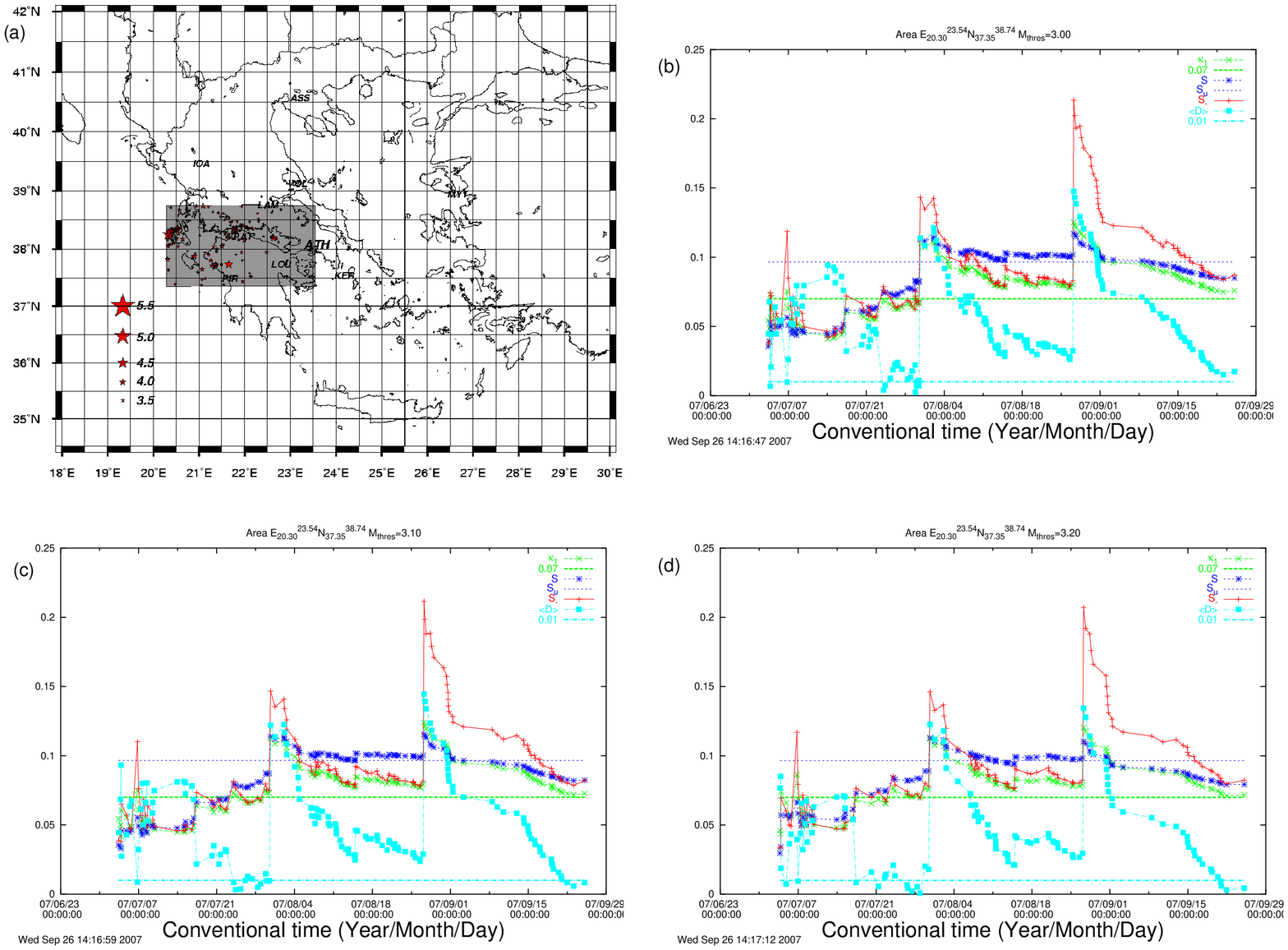}
\caption{(color online) Results of the analysis of seismicity in
the dark gray shaded area of (a) after the SES initiation on July
2, 2007. } \label{L0}
\end{figure}
\end{turnpage}

Considering the Athens observatory preliminary catalogue, the
seismic activity ($M_{thres}\ge 3.0$) that occurred in area A (see
Fig.9) after the initiation of the SES activity on July 2, 2007
(Fig.8(a)) until 03:27 UT of September 25, 2007 is shown in
Fig.\ref{A}(a). The evolution of the corresponding parameters
$\kappa_1$, $S$, $S_-$ and $\langle D \rangle$ calculated for
three magnitude thresholds, i.e.,$M_{thres}=3.0, 3.1$ and 3.2 are
shown in Figs.\ref{A}(b), (c) and (d) respectively. To investigate
the spatial invariance, the computation was repeated for several
smaller areas, three of which are shown in Figs.\ref{S1},\ref{S2}
and \ref{S3} (which are {\em different} from the areas B and C of
Fig.9) along with the evolution of the corresponding parameters.
The same was repeated for an area (see Fig.\ref{L0}) somewhat
larger than A. An inspection of all these figures, i.e.,
Figs.\ref{A} to\ref{L0}, suggests that presumably a true
coincidence has just been approached, thus being very close to the
critical point.

\section{The seismic activity that followed the most recent SES
activities at PAT and PIR}

\begin{figure}
\includegraphics{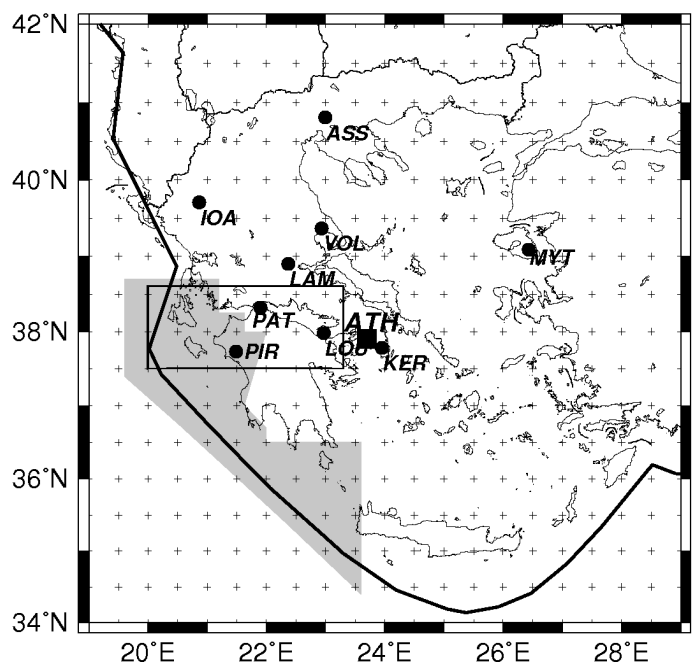}
\caption{The selectivity map of PIR (shaded area) along with the
selectivity map of PAT (rectangular area) } \label{F16}
\end{figure}

\begin{figure}
\includegraphics{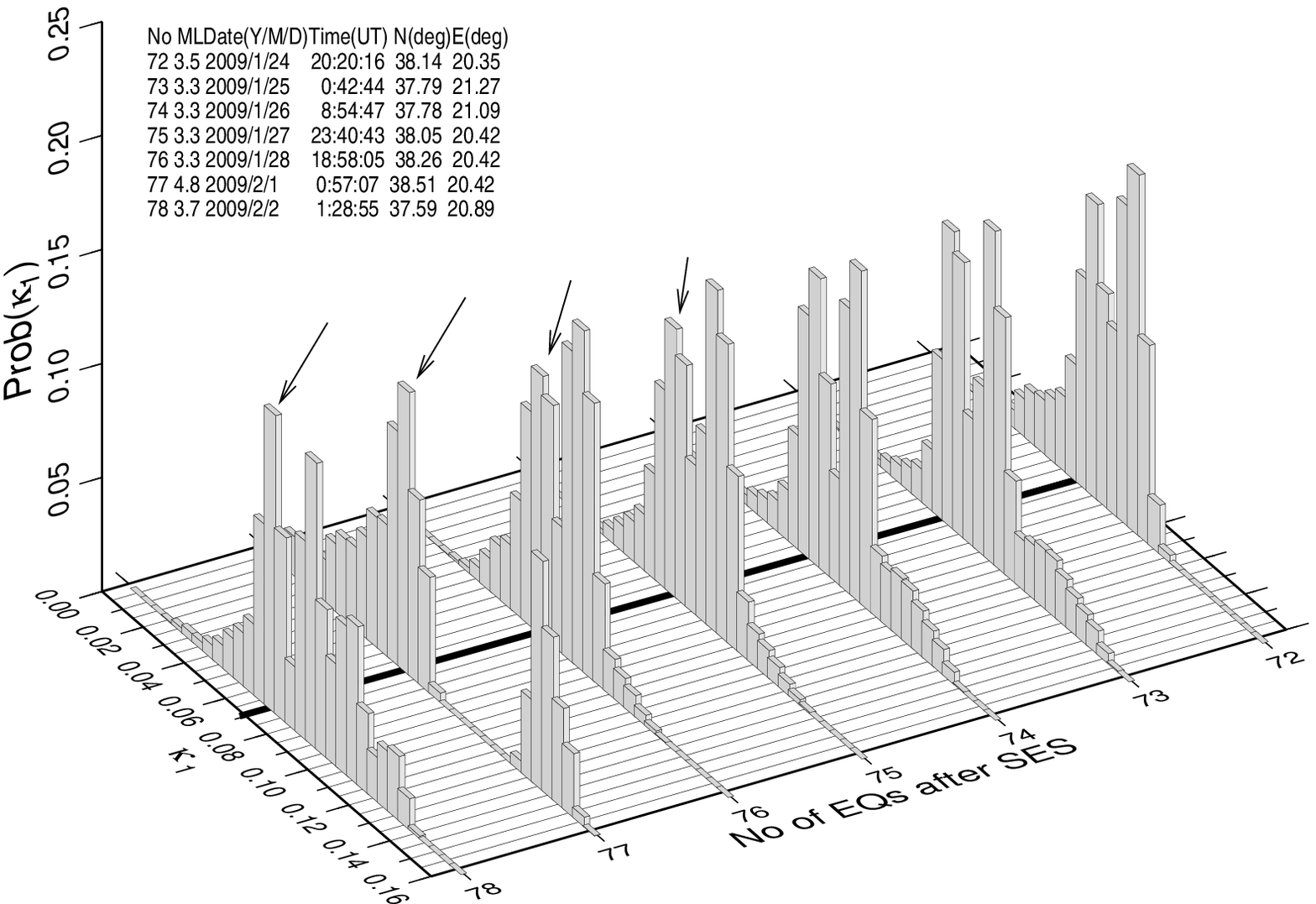}
\caption{The Prob($\kappa_1$) versus $\kappa_1$ of the seismicity
for M$_{thres}$=3.3 subsequent to the SES activity recorded at PAT
on October 9, 2008 within the selectivity map of PAT shown in Fig.
\ref{F16}. For the sake of clarity, only the last 7 events are
depicted. The arrows mark the maxima at $\kappa_1$=0.070 (see the
text) } \label{F17}
\end{figure}

\begin{figure}
\includegraphics{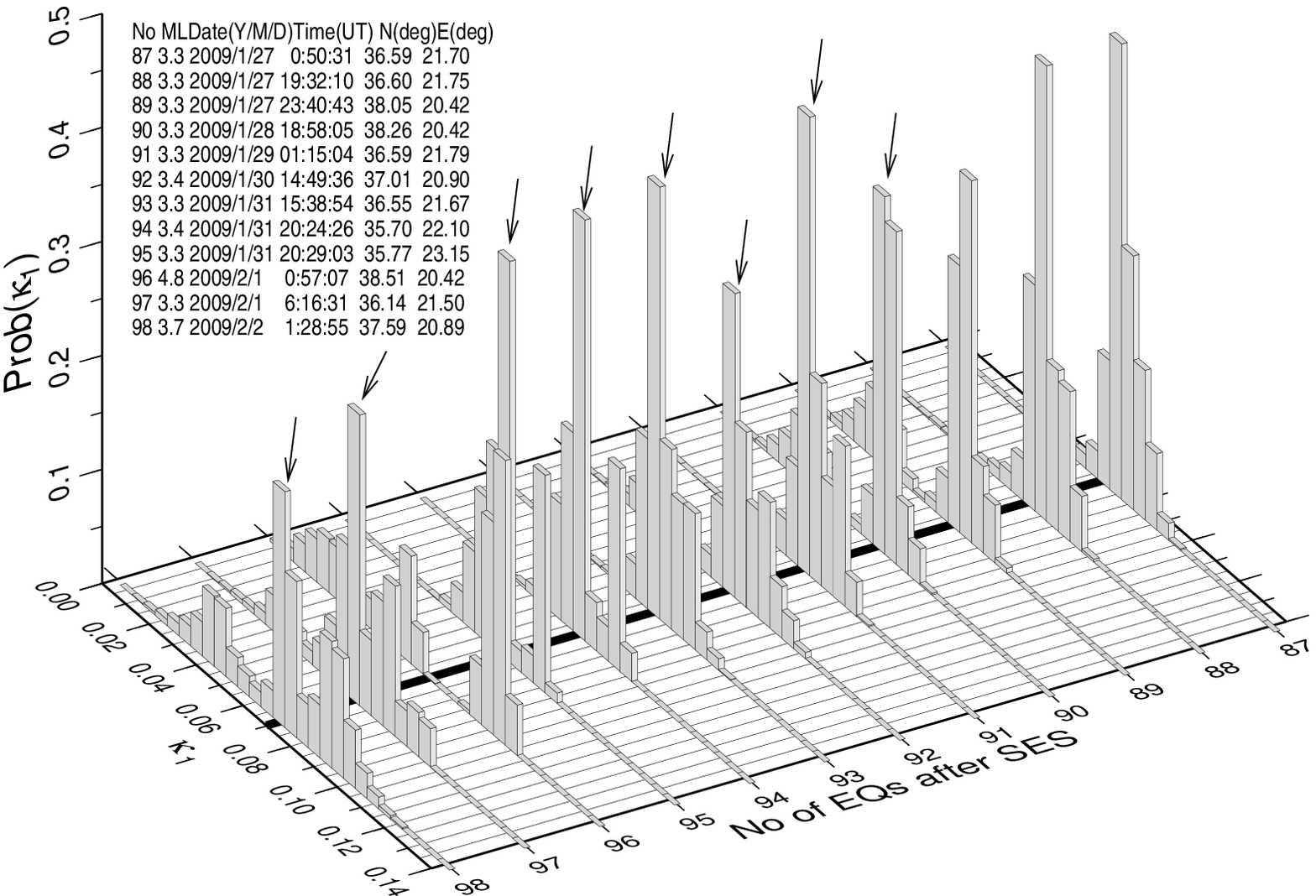}
\caption{The Prob($\kappa_1$) versus $\kappa_1$ of the seismicity
for M$_{thres}$=3.3 subsequent to the SES activity recorded at PIR
on December 12, 2008 within the selectivity map of PIR shown in
Fig. \ref{F16}.  For the sake of clarity, only the last 12 events
are depicted. The arrows mark the maxima at $\kappa_1$=0.070 (see
the text). } \label{F18}
\end{figure}

Here, we report the update results of the seismic activity that
followed the SES activity at PAT on October 9, 2008\cite{ARXIVv5}
and the SES activity at PIR on December 12, 2008\cite{CHAOS09} by
following the procedure described by Sarlis et al.\cite{JAC08}.
The subsequent seismicity of the former SES activity was studied
in the area N$_{37.5}^{38.6}$E$_{19.8}^{23.3}$ while that of the
latter in the selectivity map of PIR depicted in Fig. \ref{F16}.
The results, when considering the seismicity until early in the
morning of February 2, 2009, for magnitude threshold
M$_{thres}$=3.3, are shown in Figs. \ref{F17} and \ref{F18} for
the former and the latter SES activities at PAT and PIR,
respectively. An inspection of these figures reveals that in both
areas the probability Prob($\kappa_1$) versus $\kappa_1$
-calculated in all the possible regions of each area as described
in Ref.\cite{JAC08}- maximizes at $\kappa_1 \approx$ 0.070 upon
the occurrence of the events marked with arrows, thus probably
indicating the approach to the critical point.

{\em Note added on February 20, 2009:} Actually, at 23:16 UT on
February 16, 2009 a strong earthquake with magnitude Ms(ATH)=6.0
(i.e., M$_L$(ATH)=5.5) occurred with an epicenter at
37.1$^o$N20.8$^o$E, which clearly lies inside the selectivity map
of PIR depicted in Fig.\ref{F16}.

\begin{figure}
\includegraphics{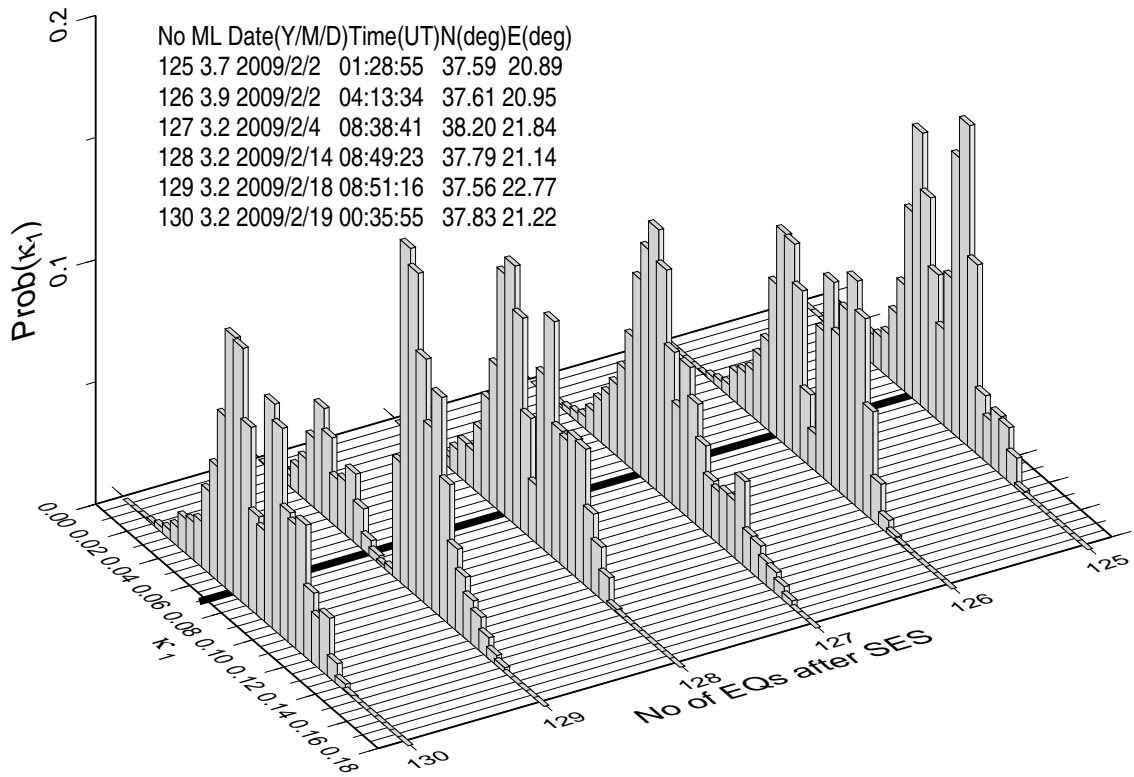}
\caption{Prob($\kappa_1$) versus $\kappa_1$ of the seismicity for
M$_{thres}$=3.2 subsequent to the SES activity recorded at PAT on
October 9, 2008 within the area
N$_{37.5}^{38.6}$E$_{19.8}^{23.3}$.  For the sake of clarity, only
the last 6 events are depicted. Prob($\kappa_1$) exhibits a
maximum at $\kappa_1$=0.070 upon the occurrence of the 3.2 event
on February 19, 2009 (see the text). } \label{F19}
\end{figure}

\begin{figure}
\includegraphics{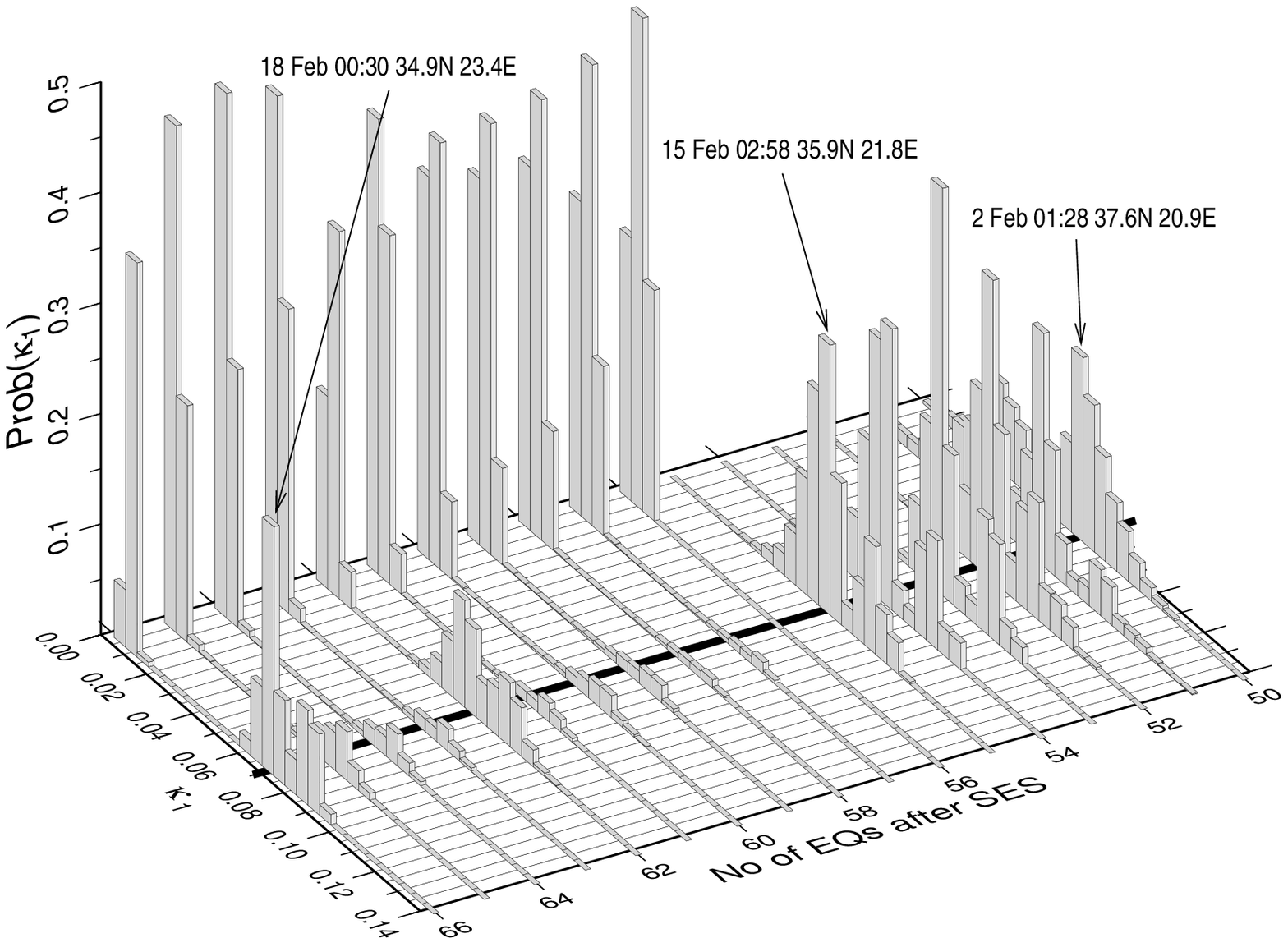}
\caption{Prob($\kappa_1$) versus $\kappa_1$ of the seismicity for
M$_{thres}$=3.4 subsequent to the SES activity recorded at PIR on
December 12, 2008 within the selectivity map of PIR shown in Fig.
\ref{F16}.  For the sake of clarity, only the last 17 events are
depicted. The arrows mark the maxima of Prob($\kappa_1$) at
$\kappa_1$=0.070 (see the text). } \label{F20}
\end{figure}

We now present the results, when considering the seismicity until
early in the morning of February 19, 2009. Figure \ref{F19},
depicts the results for M$_{thres}$=3.2 in the selectivity map of
PAT (i.e., in the area N$_{37.5}^{38.6}$E$_{19.8}^{23.3}$), which
show that the aforementioned probability Prob($\kappa_1$) versus
$\kappa_1$ maximizes at $\kappa_1 \approx$ 0.070 upon the
occurrence of a M$_L$=3.2 event at 00:35 UT on February 19, 2009
with epicenter at 37.8$^o$N21.2$^o$E. As for the results in the
selectivity map of PIR (i.e., the shaded area in Fig.\ref{F16}),
they are shown in Fig.\ref{F20} for M$_{thres}$=3.4 and reveal the
following: After the aforementioned strong earthquake on February
16, 2009, the bimodal curve Prob($\kappa_1$) versus $\kappa_1$
shows a clear local maximum at $\kappa_1 \approx$ 0.070 upon the
occurrence of the M$_L$=3.5 event at 00:30 UT on February 18, 2009
with an epicenter at 34.9$^o$N23.4$^o$E. This local maximum does
exhibit magnitude threshold invariance (since it also appears when
investigating M$_{thres}$=3.3 and M$_{thres}$=3.5).

\begin{figure}
\includegraphics{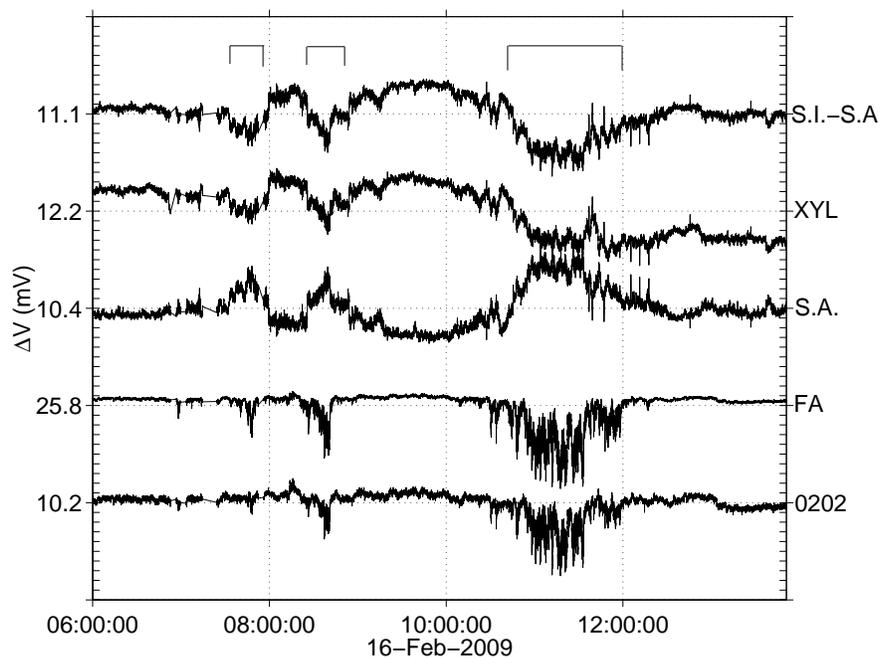}
\caption{Electrical recordings from PIR; for the sites of the
electrodes see Ref.\cite{EPAPS74}.} \label{Fig21}
\end{figure}

\begin{figure}
\includegraphics{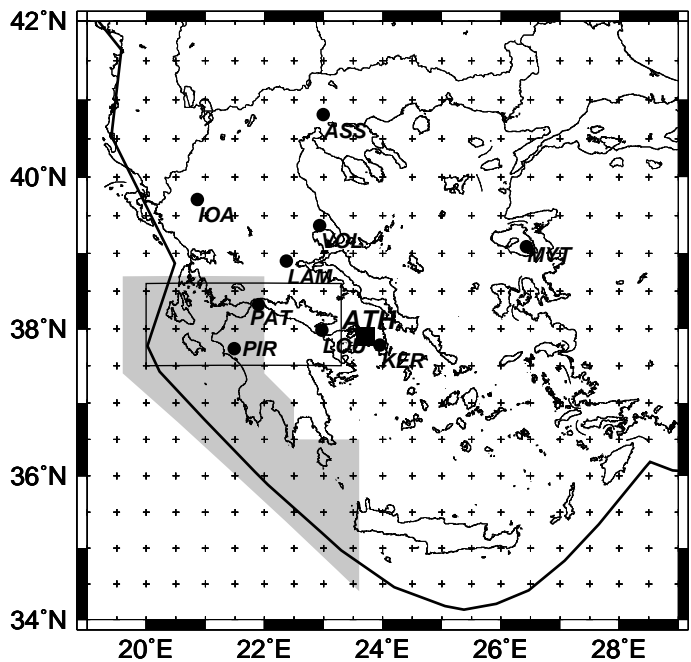}
\caption{The selectivity map of PIR when including the whole
region to the west of Hellenides (shaded area). The rectangular
area corresponds to the selectivity map of PAT.} \label{Fig22}
\end{figure}

\begin{figure}
\includegraphics{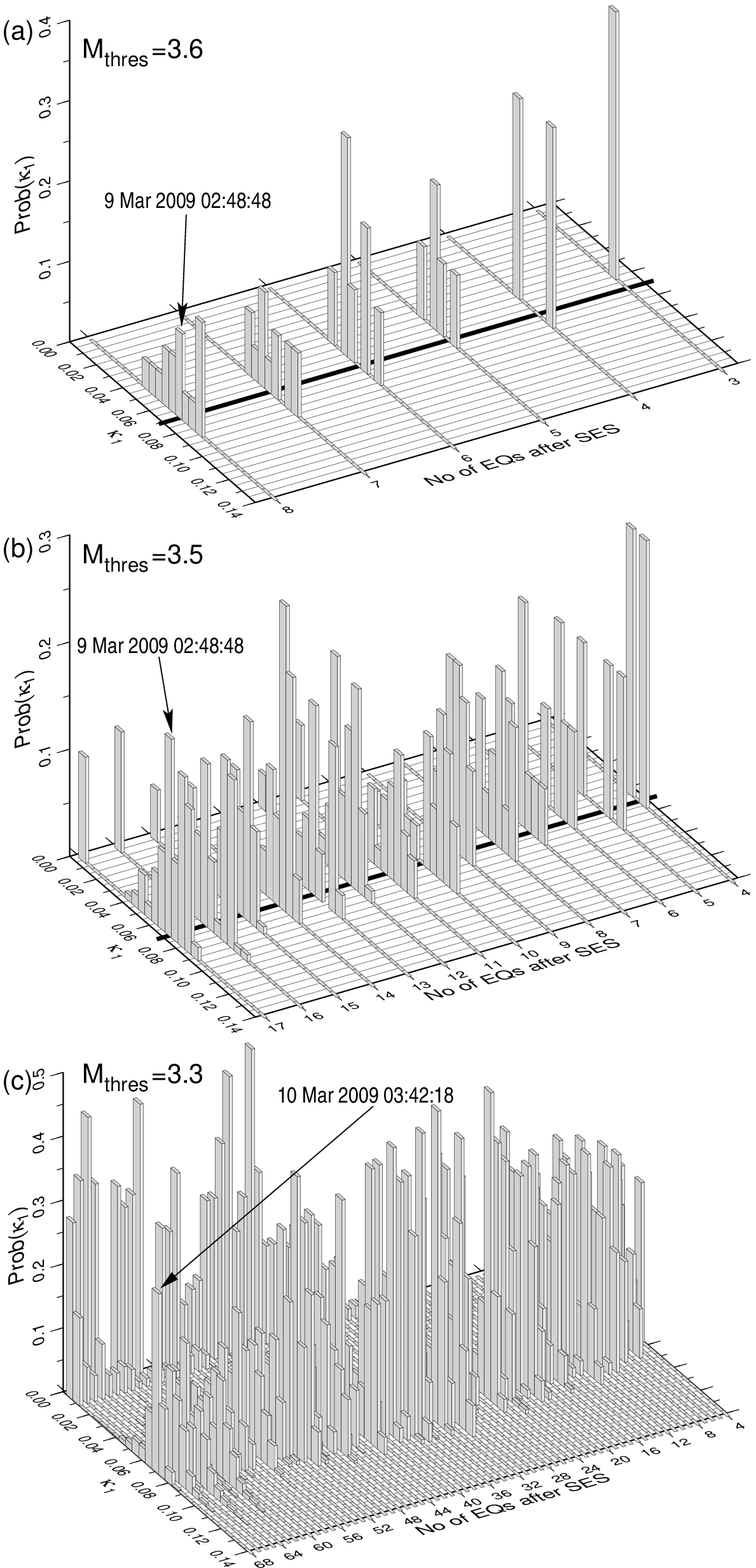}
\caption{Prob($\kappa_1$) versus $\kappa_1$ of the seismicity in
the shaded region of Fig.\ref{Fig22} for M$_{thres}$=3.6 (a), 3.5
(b) and 3.3 (c) subsequent to the electrical anomaly shown in
Fig.\ref{Fig21}. The arrows mark the maxima of Prob($\kappa_1$) at
$\kappa_1$=0.070 (see the text).} \label{Fig23}
\end{figure}

\begin{figure}
\includegraphics{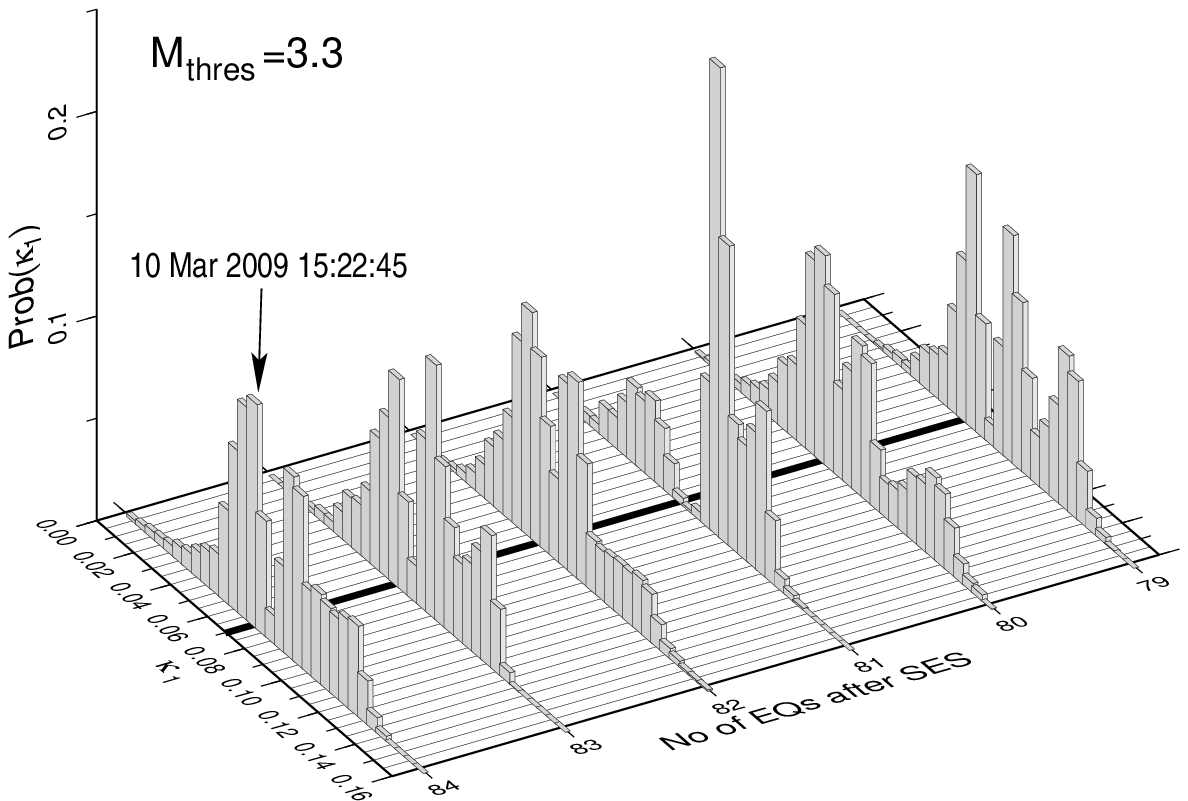}
\caption{Prob($\kappa_1$) versus $\kappa_1$ of the seismicity for
M$_{thres}$=3.3 subsequent to the SES activity recorded at PAT on
October 9, 2008 within the area
N$_{37.5}^{38.6}$E$_{19.8}^{23.3}$. For the sake of clarity, only
the last 6 events are depicted. Prob($\kappa_1$) exhibits a
maximum at $\kappa_1$=0.070 upon the occurrence of the event at
15:22 UT on March 10, 2009.} \label{Fig24}
\end{figure}

{\em Updated note on March 13, 2009:} Several hours before the
occurrence of the aforementioned magnitude 6.0 earthquake on
February 16, 2009, an electrical anomaly of significant amplitude,
see Fig.\ref{Fig21}, appeared at PIR. It basically consists of
three bay like transient changes, thus having a form essentially
different than the SES activities that preceded the major
earthquakes on February 14, 2008 and June 8, 2008. This electrical
anomaly could be in principle attributed to the 6.0 earthquake on
February 16, 2009 that occurred almost 11 hours later, but this
earthquake was also preceded by the SES activity at PIR on
December 12, 2008. Alternatively, this anomaly might be related to
a new impending earthquake. For this reason, a natural time
analysis of the seismicity subsequent to the electrical anomaly of
February 16, 2009 was carried out. In this calculation we
considered the events that occurred in the PIR selectivity map
shown in Fig.\ref{Fig22} (cf. this, which is slightly different
than that depicted in Fig.16, has been drawn by taking into
account the whole region to the west of the Hellenides). The
results of the analysis shown in Fig.\ref{Fig23} indicate that for
M$_{thres}$=3.6 and 3.5 (see a and b, respectively) the
probability Prob($\kappa_1$) maximizes at $\kappa_1 \approx$ 0.070
upon the occurrence of a M$_L$=3.6 event at 02:48 UT on March 9,
 2009, while for M$_{thres}$=3.3 the maximization occurs upon the
occurrence of a M$_L$=3.3 event on March 10, 2009. Interestingly,
a similar analysis (subsequent to the SES
 activity at PAT) in the selectivity map of PAT, i.e., in the area
 N$_{37.5}^{38.6}$E$_{19.8}^{23.3}$, reveals that a maximization
 of Prob($\kappa_1$) at $\kappa_1 \approx$ 0.070 also occurs on
 March 10, 2009 (see Fig.\ref{Fig24}).
 We are currently investigating the spatial and magnitude
 threshold invariance of the aforementioned maxima to examine
 whether they actually point to an approach to the critical
 point.


\begin{thebibliography}{40}
\bibitem[{mag()}]{magest}
\eprint{In the natural time analysis of the seismicity we study
the evolution
  of the pair ($\chi_k$, $Q_k$), where $Q_k$ is proportional to the seismic
  moment $M_0$. Following Refs.[1-4,9], for the reasons of this calculation the
  formulae $M_0=10^{M_L+10.5}$(Nm), for $M_L \le 3.6$ and
  $M_0=10^{1.5M_L+8.7}$(Nm), for $3.6 < M_L \le 5.0$ proposed by the Global
  Seismological Services were used. Then, the normalized power spectrum is
  given[1-4,9] by \[ \Pi (\phi) \equiv \left| \sum_{k=1}^N p_k \exp\left(
  \text{i} 2 \pi \phi \chi_k \right) \right|^2, \] where $p_k=\left( M_0
  \right)_k/\sum_{n=1}^N \left( M_0 \right)_n$.}
\end{thebibliography}

\newpage
\newpage
\newpage
\newpage

\newpage

\newpage

\newpage

\begin{table*}
\caption{The catalogue ($M_{thres}=3.2$) of the Institute of
Geodynamics of the National Observatory of Athens (GI-NOA)
(available from {\tt
http://www.gein.noa.gr/services/monthly-list.html} on February 5,
2007) for the   Large area (area A) during the following period:
From the initiation of the SES activity almost at 17:00 UT on
November 14, 2006 (Fig.1 of the main text) until the occurrence of
the 5.8-units earthquake on February 3, 2007. Note that
Ms(ATH)=M+0.5, where M stands for the local magnitude ML or the
``duration'' magnitude MD. Two events reported by GI-NOA: at
17:32:23.7 on January 28, 2007, and at 5:11:06.5 on January 30,
2007, were omitted since they could not be independently confirmed
by other seismic networks operating at that time.} \label{tab60}
\begin{ruledtabular}
\begin{tabular}{ccccccccc}
$N$ & Year & Month & Day & UT  &
Lat.($^o$N) & Lon.($^o$E) & depth(km) & M  \\
\hline
     1 &   2006 &  11 &  16 &  4:20:40.0 &  35.66 &  23.26 &     54 &  3.5   \\
     2 &   2006 &  11 &  16 & 12:30:50.3 &  35.44 &  23.42 &      5 &  3.7   \\
     3 &   2006 &  11 &  16 & 18:50:02.5 &  36.97 &  23.11 &     25 &  3.7   \\
     4 &   2006 &  11 &  16 & 21:53:12.5 &  35.72 &  24.33 &     10 &  3.4   \\
     5 &   2006 &  11 &  17 & 16:31:26.4 &  36.69 &  23.32 &     24 &  3.2   \\
     6 &   2006 &  11 &  18 &  3:48:58.3 &  34.79 &  24.07 &     16 &  3.6   \\
     7 &   2006 &  11 &  18 & 21:51:21.8 &  34.78 &  24.36 &     34 &  3.2   \\
     8 &   2006 &  11 &  19 &  9:31:09.1 &  35.73 &  21.95 &     19 &  3.6   \\
     9 &   2006 &  11 &  19 & 10:54:52.3 &  35.99 &  23.65 &     10 &  3.2   \\
    10 &   2006 &  11 &  19 & 14:11:45.4 &  35.75 &  22.55 &     24 &  3.3   \\
    11 &   2006 &  11 &  21 &  5:57:05.5 &  34.81 &  24.28 &     54 &  3.3   \\
    12 &   2006 &  11 &  21 &  8:42:06.0 &  37.42 &  21.96 &     19 &  3.2   \\
    13 &   2006 &  11 &  21 & 12:50:34.0 &  35.45 &  23.90 &     10 &  3.3   \\
    14 &   2006 &  11 &  24 & 17:16:58.4 &  37.16 &  21.58 &     12 &  4.1   \\
    15 &   2006 &  11 &  24 & 17:42:56.3 &  35.60 &  23.50 &      7 &  3.3   \\
    16 &   2006 &  11 &  24 & 20:26:02.9 &  37.12 &  21.56 &     11 &  3.5   \\
    17 &   2006 &  11 &  24 & 21:41:44.1 &  37.21 &  21.63 &      8 &  3.3   \\
    18 &   2006 &  11 &  25 &  3:14:31.2 &  34.94 &  23.91 &      5 &  3.2   \\
    19 &   2006 &  11 &  25 &  4:55:01.3 &  37.15 &  21.86 &      9 &  3.3   \\
    20 &   2006 &  11 &  25 &  9:30:18.1 &  36.75 &  21.73 &     12 &  3.6   \\
    21 &   2006 &  11 &  26 & 17:58:05.5 &  35.45 &  23.30 &     11 &  4.1   \\
    22 &   2006 &  11 &  26 & 19:45:54.7 &  35.40 &  23.27 &      2 &  3.3   \\
    23 &   2006 &  11 &  26 & 23:57:49.1 &  35.47 &  23.36 &      5 &  3.4   \\
    24 &   2006 &  11 &  27 &  3:12:49.8 &  35.42 &  23.31 &      9 &  3.3   \\
    25 &   2006 &  11 &  27 & 11:58:03.7 &  35.53 &  23.43 &     24 &  3.2   \\
    26 &   2006 &  11 &  28 &  9:07:38.6 &  36.07 &  22.35 &     10 &  3.3   \\
    27 &   2006 &  11 &  28 & 15:07:43.1 &  36.06 &  22.42 &     43 &  3.6   \\
    28 &   2006 &  11 &  28 & 15:42:59.4 &  35.71 &  22.11 &     10 &  3.4   \\
    29 &   2006 &  11 &  28 & 22:54:24.0 &  37.06 &  21.51 &     12 &  3.2   \\
    30 &   2006 &  11 &  29 & 14:11:02.8 &  35.97 &  23.11 &     19 &  3.5   \\
    31 &   2006 &  11 &  30 & 23:30:58.6 &  35.43 &  23.32 &      3 &  3.2   \\
    32 &   2006 &  12 &   1 &  2:37:02.1 &  35.58 &  23.60 &     10 &  3.5   \\
    \end{tabular}
\end{ruledtabular}
\end{table*}

\addtocounter{table}{-1}
\begin{table*}
\caption{Continued}
\begin{ruledtabular}

\begin{tabular}{ccccccccc}
$N$ & Year & Month & Day & UT  &
Lat.($^o$N) & Lon.($^o$E) & depth(km) & M  \\
\hline
33 &   2006 &  12 &   1 & 19:57:55.4 &  37.07 &  21.47 &      3 &  3.2   \\
    34 &   2006 &  12 &   1 & 22:49:35.6 &  35.72 &  22.52 &    114 &  3.7   \\
    35 &   2006 &  12 &   2 &  0:29:20.8 &  34.74 &  22.87 &     45 &  3.2   \\
36 &   2006 &  12 &   2 & 10:30:35.3 &  37.42 &  22.04 &      4 &  3.3   \\
    37 &   2006 &  12 &   3 &  1:53:22.4 &  36.72 &  21.81 &     24 &  3.5   \\
    38 &   2006 &  12 &   3 &  2:37:36.7 &  36.40 &  21.66 &     10 &  3.2   \\
    39 &   2006 &  12 &   4 & 19:32:24.0 &  34.99 &  23.44 &     14 &  3.4   \\
    40 &   2006 &  12 &   7 & 22:22:49.4 &  35.73 &  23.17 &     32 &  3.2   \\
    41 &   2006 &  12 &   8 & 10:54:40.0 &  34.95 &  23.42 &      5 &  3.3   \\
    42 &   2006 &  12 &   8 & 12:48:15.5 &  37.02 &  21.93 &     10 &  3.2   \\
    43 &   2006 &  12 &   8 & 19:57:27.2 &  35.06 &  23.50 &      7 &  3.6   \\
    44 &   2006 &  12 &   9 & 22:45:56.5 &  35.89 &  23.38 &     24 &  3.4   \\
    45 &   2006 &  12 &  10 &  3:29:46.1 &  34.84 &  24.48 &     10 &  3.2   \\
    46 &   2006 &  12 &  10 & 12:41:27.6 &  35.80 &  23.04 &     59 &  3.2   \\
    47 &   2006 &  12 &  11 &  3:36:22.0 &  35.74 &  24.08 &     16 &  3.4   \\
    48 &   2006 &  12 &  11 & 19:11:29.4 &  35.93 &  23.40 &     15 &  3.2   \\
    49 &   2006 &  12 &  12 &  3:40:55.0 &  35.54 &  22.81 &     17 &  3.7   \\
    50 &   2006 &  12 &  13 &  9:45:40.0 &  37.25 &  21.94 &     20 &  3.2   \\
    51 &   2006 &  12 &  14 & 16:44:02.0 &  35.95 &  23.46 &     10 &  3.3   \\
    52 &   2006 &  12 &  15 &  4:40:12.5 &  35.01 &  23.37 &     80 &  3.4   \\
    53 &   2006 &  12 &  15 & 10:23:32.1 &  36.09 &  22.25 &     45 &  3.7   \\
    54 &   2006 &  12 &  16 & 10:43:15.4 &  34.85 &  24.33 &      6 &  3.6   \\
    55 &   2006 &  12 &  16 & 15:23:33.2 &  34.92 &  23.45 &      5 &  3.7   \\
    56 &   2006 &  12 &  17 &  4:44:38.7 &  34.73 &  24.00 &     30 &  3.5   \\
    57 &   2006 &  12 &  17 &  4:44:42.7 &  35.15 &  24.12 &     59 &  3.5   \\
    58 &   2006 &  12 &  17 & 10:20:49.7 &  36.06 &  21.76 &     37 &  3.3   \\
    59 &   2006 &  12 &  17 & 14:13:23.4 &  36.21 &  21.70 &     10 &  3.4   \\
    60 &   2006 &  12 &  17 & 20: 0:20.6 &  34.84 &  24.21 &     34 &  4.0   \\
    61 &   2006 &  12 &  17 & 22:38:37.5 &  36.65 &  21.00 &     10 &  3.2   \\
    62 &   2006 &  12 &  20 &  4:34:09.2 &  37.48 &  21.50 &     10 &  3.2   \\
    63 &   2006 &  12 &  20 &  4:45:15.0 &  35.42 &  21.43 &     10 &  3.5   \\
    64 &   2006 &  12 &  20 & 19:27:32.5 &  34.60 &  23.79 &     11 &  3.8   \\
    65 &   2006 &  12 &  20 & 20:17:02.7 &  36.68 &  21.50 &      4 &  3.2   \\
    66 &   2006 &  12 &  24 &  6:58:02.3 &  34.94 &  24.03 &     10 &  3.6   \\
    67 &   2006 &  12 &  24 &  7:12:11.4 &  36.29 &  22.20 &     24 &  3.8   \\
    68 &   2006 &  12 &  25 & 14:08:59.4 &  34.83 &  22.66 &     39 &  4.4   \\
    69 &   2006 &  12 &  25 & 14:15:50.1 &  34.99 &  23.04 &     39 &  4.0   \\
    70 &   2006 &  12 &  25 & 14:18:50.7 &  35.09 &  23.06 &     27 &  4.0   \\
    71 &   2006 &  12 &  25 & 14:22:30.3 &  35.01 &  22.95 &     18 &  3.9   \\
    72 &   2006 &  12 &  25 & 14:57:00.5 &  35.04 &  22.84 &     14 &  4.1   \\
    73 &   2006 &  12 &  25 & 17:29:17.9 &  35.92 &  23.62 &      5 &  3.2   \\
    74 &   2006 &  12 &  25 & 18:13:44.5 &  35.09 &  23.29 &     10 &  3.7   \\
    75 &   2006 &  12 &  28 &  6:32:19.4 &  37.53 &  21.81 &     10 &  3.2   \\
    76 &   2006 &  12 &  29 &  2:00:54.9 &  36.77 &  21.81 &      5 &  3.3   \\
    77 &   2006 &  12 &  30 & 22:51:38.0 &  35.39 &  23.34 &     13 &  3.2   \\
\end{tabular}
\end{ruledtabular}
\end{table*}

\addtocounter{table}{-1}
\begin{table*}
\caption{Continued}
\begin{ruledtabular}

\begin{tabular}{ccccccccc}
$N$ & Year & Month & Day & UT  &
Lat.($^o$N) & Lon.($^o$E) & depth(km) & M  \\
\hline
       78 &   2006 &  12 &  31 &  7:00:51.9 &  35.85 &  22.08 &     10 &  3.2   \\
    79 &   2006 &  12 &  31 & 18: 6:27.5 &  35.14 &  22.76 &     12 &  3.7   \\
    80 &   2007 &   1 &   1 & 15:27:48.7 &  34.72 &  24.12 &     23 &  3.5   \\
    81 &   2007 &   1 &   3 & 15:04:04.9 &  35.37 &  23.29 &     10 &  3.4   \\
    82 &   2007 &   1 &   3 & 18:18:34.5 &  36.54 &  21.74 &     26 &  3.8   \\
    83 &   2007 &   1 &   4 &  7:57:08.3 &  37.37 &  21.56 &     15 &  3.3   \\
    84 &   2007 &   1 &   4 & 14:55:24.6 &  36.29 &  21.92 &     10 &  3.2   \\
    85 &   2007 &   1 &   4 & 17:42:54.7 &  34.91 &  23.61 &     45 &  3.3   \\
    86 &   2007 &   1 &   4 & 20:41:40.9 &  34.89 &  24.13 &     48 &  3.2   \\
    87 &   2007 &   1 &   4 & 22:47:39.1 &  36.96 &  21.07 &     25 &  3.2   \\
    88 &   2007 &   1 &   7 & 14:07:11.5 &  37.10 &  21.93 &     26 &  3.2   \\
    89 &   2007 &   1 &   8 & 16:13:02.8 &  35.10 &  23.03 &     10 &  3.4   \\
    90 &   2007 &   1 &   9 &  9:55:29.2 &  36.70 &  21.55 &     31 &  3.4   \\
    91 &   2007 &   1 &   9 & 15:54:47.3 &  35.89 &  23.61 &      3 &  3.2   \\
    92 &   2007 &   1 &   9 & 23:09:20.0 &  36.21 &  22.70 &    116 &  3.6   \\
    93 &   2007 &   1 &  11 &  5:50:39.5 &  35.02 &  22.48 &     38 &  3.8   \\
    94 &   2007 &   1 &  13 & 11:12:43.3 &  35.48 &  23.51 &      8 &  3.4   \\
    95 &   2007 &   1 &  14 &  9:09:23.0 &  35.30 &  23.38 &      3 &  3.8   \\
    96 &   2007 &   1 &  14 & 16:43:01.6 &  35.06 &  23.20 &     85 &  4.1   \\
    97 &   2007 &   1 &  15 &  0:55:18.8 &  37.47 &  21.07 &     15 &  3.2   \\
    98 &   2007 &   1 &  15 &  3:13:45.8 &  36.06 &  22.40 &     13 &  3.4   \\
    99 &   2007 &   1 &  15 &  6:56:47.0 &  35.42 &  23.56 &     33 &  3.3   \\
   100 &   2007 &   1 &  15 & 17:50:48.7 &  36.56 &  21.65 &      5 &  3.4   \\
   101 &   2007 &   1 &  17 &  1:52:18.8 &  36.20 &  21.58 &     34 &  3.7   \\
   102 &   2007 &   1 &  17 &  3:22:53.9 &  35.34 &  23.53 &     12 &  3.7   \\
   103 &   2007 &   1 &  18 & 22:25:23.0 &  34.84 &  22.67 &     39 &  4.7   \\
   104 &   2007 &   1 &  18 & 23:31:32.7 &  34.76 &  21.36 &     10 &  3.4   \\
   105 &   2007 &   1 &  19 &  8:42:22.8 &  34.72 &  22.57 &     38 &  4.4   \\
   106 &   2007 &   1 &  20 &  7:05:37.0 &  34.95 &  24.09 &     19 &  3.5   \\
   107 &   2007 &   1 &  20 &  7:24:24.6 &  34.86 &  23.96 &     19 &  3.4   \\
   108 &   2007 &   1 &  20 & 10:24:55.4 &  36.88 &  22.10 &     10 &  3.2   \\
   109 &   2007 &   1 &  20 & 11:57:23.7 &  36.69 &  22.47 &     10 &  3.5   \\
   110 &   2007 &   1 &  20 & 23:50:04.1 &  34.93 &  24.40 &     28 &  3.3   \\
   111 &   2007 &   1 &  21 & 17:52:34.9 &  35.46 &  23.46 &     11 &  3.4   \\
   112 &   2007 &   1 &  25 & 10:24:47.3 &  35.33 &  23.40 &      2 &  3.3   \\
   113 &   2007 &   1 &  25 & 21:09:45.7 &  35.09 &  23.25 &     22 &  4.3   \\
   114 &   2007 &   1 &  28 & 15:19:23.6 &  36.29 &  23.17 &     25 &  3.4   \\
   115 &   2007 &   1 &  28 & 19:12:16.4 &  36.95 &  21.13 &     10 &  3.4   \\
   116 &   2007 &   1 &  30 &  5:11:08.9 &  35.94 &  23.41 &     28 &  3.3   \\
   117 &   2007 &   1 &  30 & 10:38:42.7 &  36.98 &  21.10 &     12 &  3.5   \\
   118 &   2007 &   1 &  30 & 21:11:43.0 &  36.13 &  22.17 &     30 &  3.3   \\
   119 &   2007 &   1 &  31 &  4:56:59.1 &  35.49 &  22.78 &     36 &  3.4   \\
   120 &   2007 &   1 &  31 & 14:45:09.7 &  34.67 &  22.42 &     29 &  3.6   \\
   121 &   2007 &   1 &  31 & 18:40:54.1 &  36.26 &  22.48 &     23 &  3.6   \\
   122 &   2007 &   2 &   1 & 16:45:35.3 &  34.69 &  22.46 &     21 &  3.9   \\
   123 &   2007 &   2 &   1 & 21:06:54.7 &  36.06 &  21.62 &     42 &  3.2   \\
   124 &   2007 &   2 &   2 &  6:39:05.6 &  35.21 &  23.25 &     33 &  3.3   \\
   125 &   2007 &   2 &   2 & 13:27:53.1 &  37.22 &  21.63 &      4 &  3.4
\end{tabular}
\end{ruledtabular}
\end{table*}
\newpage

\end{document}